\documentclass[11pt,draftcls,onecolumn]{IEEEtran}
\usepackage{amsfonts,bm,amsmath,comment,color,cite,amssymb,subfigure,multirow}
\usepackage[]{graphicx}
\usepackage[linesnumbered,ruled]{algorithm2e}

\def\bgam{{\boldsymbol{\gamma}}}
\def\bTheta{{\boldsymbol{\Theta}}}

\def\blam{{\boldsymbol{\lambda}}}

\def\bzero{{\mathbf{0}}}

\def\c1{{\textcircled{a}}}
\def\ba{{\mathbf{a}}}
\def\bb{{\mathbf{b}}}

\def\bd{{\mathbf{d}}}
\def\be{{\mathbf{e}}}

\def\bm{{\mathbf{m}}}
\def\bn{{\mathbf{n}}}

\def\bq{{\mathbf{q}}}

\def\bs{{\mathbf{s}}}
\def\bt{{\mathbf{t}}}
\def\bu{{\mathbf{u}}}
\def\bv{{\mathbf{v}}}
\def\bw{{\mathbf{w}}}
\def\bx{{\mathbf{x}}}
\def\by{{\mathbf{y}}}
\def\bz{{\mathbf{z}}}
\def\bA{{\mathbf{A}}}
\def\bB{{\mathbf{B}}}
\def\bC{{\mathbf{C}}}
\def\bD{{\mathbf{D}}}

\def\bG{{\mathbf{G}}}
\def\bH{{\mathbf{H}}}
\def\bI{{\mathbf{I}}}

\def\bN{{\mathbf{N}}}
\def\bM{{\mathbf{M}}}

\def\bQ{{\mathbf{Q}}}
\def\bR{{\mathbf{R}}}
\def\bS{{\mathbf{S}}}
\def\bT{{\mathbf{T}}}
\def\bU{{\mathbf{U}}}
\def\bV{{\mathbf{V}}}

\def\bX{{\mathbf{X}}}
\def\bY{{\mathbf{Y}}}

\def\bzero{{\mathbf{0}}}

\def\tr{{\textrm{tr}}}

\def\NT{{N_\textrm{T}}}
\def\NR{{N_\textrm{R}}}
\def\txT{{\textrm{T}}}
\def\txR{{\textrm{R}}}

\def\HH{{\dagger}}
\renewcommand\Re{{\textrm{Re}}}
\renewcommand\vec{{\textrm{vec}}}

\begin{document}
%
\title{MIMO Multifunction RF Systems: Detection Performance and Waveform Design}
\author{Bo~Tang,~
             Petre~Stoica,~\IEEEmembership{Fellow,~IEEE}
\thanks{The work of Bo Tang was supported in part by the National Natural Science Foundation of China under Grants  62171450 and 61671453,
and Anhui Provincial Natural Science Foundation under Grant 2108085J30. The work of Petre Stoica was supported by the Swedish Research Council (VR Grants 2017-04610 and 2016-06079).}
\thanks{Bo Tang is with the College of Electronic Engineering, National University of Defense Technology, Hefei 230037, China
        (Email: tangbo06@gmail.com). }
\thanks{Petre Stoica is with the Department of Information Technology, Uppsala University, 75105 Uppsala, Sweden (Email: ps@it.uu.se).}
\thanks{This work was presented in part at the 2021 CIE International Conference on Radar, Haikou, China \cite{Tang2021radarconf}.}
}


\maketitle

\begin{abstract}
This paper studies the detection performance of a multiple-input-multiple-output (MIMO) multifunction radio frequency (MFRF) system, which simultaneously supports radar, communication, and jamming. We show that the detection performance of the MIMO MFRF system improves as the transmit signal-to-interference-plus-noise-ratio (SINR) increases. To analyze the achievable SINR of the system, we formulate an SINR maximization problem under the communication and jamming functionality constraint as well as a transmit energy constraint. We derive a closed-form solution of this optimization problem for energy-constrained waveforms and present a detailed analysis of the achievable SINR. Moreover, we analyze the SINR for systems transmitting constant-modulus waveforms,  which are often used in practice. We propose an efficient constant-modulus waveform design algorithm to maximize the SINR. Numerical results demonstrate the capability of a MIMO array to provide multiple functions, and also show the tradeoff between radar detection and the communication/jamming functionality.
\end{abstract}
%


\newtheorem{Def}{Definition}
\newtheorem{lemma}{Lemma}
\newtheorem{theorem}{Theorem}
\newtheorem{Prop}{Proposition}

\section{Introduction}
Multifunction radio frequency (MFRF) systems integrate several functions, including radar, communications and electronic warfare (EW), through shared aperture, radio frequency (RF) components, and signal processing hardware \cite{tavik2005advanced}. Compared with traditional RF systems, which lack the level of integration of MFRF systems, the latter systems have many advantages, some of which can be summarized as follows: 1)  the number  of antennas and the radar cross section of  the platform is reduced; 2) the weight of the platform is decreased and the maneuverability is enhanced; 3) the mutual interference among nearby systems (e.g., radar and communication systems) is minimized and the spectral compatibility is improved \cite{tavik2005advanced}. Due to their great potential, the MFRF systems have received considerable attention from both academia and industry \cite{tavik2005advanced,moo2018overview,hemmi1996ASAP,ouacha2010MAESA}.

\subsection{Background and Related Work}
The idea of using a single RF system to provide multiple functions dates back to 1960s \cite{Mealey1963RadarCom}. In \cite{Mealey1963RadarCom}, the authors proposed a radar communication system, which can measure the target range as well as transmit information to a transponder. However, the number of data bits conveyed in a pulse group was low.  In \cite{tavik2005advanced,moo2018overview,hemmi1996ASAP,ouacha2010MAESA}, wideband antenna arrays were developed to support radar, communication, and EW simultaneously. The aperture of the developed antenna was split into several parts, and each part was configured to support a single RF function (radar, communication, or EW). However, the split aperture resulted in a reduced equivalent isotropic radiated power and thus a poorer performance (e.g., a shorter range for radar detection). Moreover, the emission of multiple waveforms in separated frequency bands consumed more spectral resources and lead to spectral congestion.
To fully utilize the antenna aperture and improve the spectral compatibility, many studies have been devoted to designing a shared waveform to support multiple functions (more precisely, a dual function). In \cite{barrenechea2007fmcw}, the authors used amplitude modulation (AM) to encode data onto a frequency modulated continuous wave (FMCW) radar signal. In \cite{Sturm2011OFDM,Liu2017OFDM}, orthogonal frequency division multiplexing (OFDM) signals were proposed for simultaneous radar sensing and data communication. However, OFDM signals suffer from the envelope variation problem, which might cause non-linear effects in the transmitters. In addition to the AM-FMCW signals and the OFDM signals, there are also attempts of modulating a chirp signal (i.e., a linear frequency modulated signal) with a communication signal (see, e.g., \cite{chen2011LFMMSK,nowak2017mixed}). Compared with the OFDM signals, the constant-modulus  mixed-modulated signals allow the radio frequency amplifier to operate at maximum efficiency. However, such signals have high autocorrelation sidelobes, meaning that weak target reflections might be obscured by returns from strong targets. In \cite{Ciuonzo2015Intrapulse}, an intrapulse radar-embedded communication scheme was proposed to covertly deliver the information bits. Through the elaborate design of communication waveforms (e.g., by multi-objective optimization algorithms), the proposed scheme can achieve a low probability of interception by eavesdroppers.

More recently, there has been a growing interest in using multiple-input-multiple-output (MIMO) arrays to realize dual-function radar-communication systems
\cite{Hassanien2016Embedding,Wang2019SparseArray,McCormick2017Simultaneous,Liu2018DFRC,Zhang2021Overview,Liu2020Beamforming}. Compared with conventional phased-array systems, the waveform diversity provided by MIMO systems enables emitting different  signals in different directions. In \cite{Hassanien2016Embedding}, the authors proposed to deliver information bits to the users  by properly designing the transmit beampattern of the MIMO array. In \cite{McCormick2017Simultaneous}, the authors suggested a waveform design algorithm for simultaneous transmission of radar and communication signals using a shared MIMO array.  Different from  \cite{McCormick2017Simultaneous}, the authors in \cite{Liu2018DFRC} considered the design of multiple waveforms, with the aim of matching a desired beampattern for radar sensing and synthesizing modulated signals for data delivery (see also \cite{Tang2020SAM,Shi2020DFRC} for similar studies). To improve the radar detection performance in clutter and ensure communication capabilities,  the authors in \cite{Tsinos2020JointDesign} proposed joint design of the transmit waveform and receive filter for a MIMO array. In \cite{li2022dual}, the authors showed that with the aid of intelligent reflecting surface, it was possible to detect a non-line-of-sight target as well as communicate with multiple users simultaneously.

\subsection{Contributions}
Note that the studies reviewed above mainly focus on realizing a dual function (typically, radar detection  and communication). In this paper, we show that it is possible to realize an MFRF system based on a MIMO array. We call this system, which simultaneously supports radar, communication, and jamming, a MIMO MFRF system.  Such a system is useful if we would like to detect a target of interest, and communicate with multiple friendly targets. To communicate covertly, we may also want to transmit jamming signals toward the hostile targets (which can be eavesdroppers trying to intercept the communication signals).

The main focus of our study is on analyzing the target detection performance of a MIMO MFRF system. To this end, we formulate a hypothesis testing problem and derive the detection probability of the associated Neyman-Pearson (NP) detector. We show that the detection probability is a monotonically increasing function of the transmit signal-to-interference-plus-noise ratio (SINR). Consequently, we formulate an SINR maximization problem under the communication and the jamming functionality constraints. Through tackling the formulated optimization problem, we analyze the achievable SINR of the system and design the associated waveforms. Results are provided to demonstrate the impact of supporting multiple functions on the detection performance of MFRF systems.

\subsection{Organization}
The rest of this paper is organized as follows. Section \ref{sec:SigModel} presents the signal model and formulates the problem.
Section \ref{sec:OptimalSolution} derives the optimal waveforms maximizing the SINR under the multifunction constraint. Section \ref{sec:structured} studies the achievable SINR of MIMO MFRF systems in detail. Section \ref{sec:PAPR} derives an efficient algorithm to design waveforms under a low peak-to-average-power ratio (PAPR) constraint. Section \ref{sec:Discussion} analyzes the computational complexity of the proposed algorithm and the performance of the system. Section \ref{sec:NumericalExamples} provides numerical examples to demonstrate the performance of MIMO MFRF systems. Finally, we draw the conclusions in Section \ref{sec:Conclusion}.

\subsection{Notation}
Matrices are denoted by bold uppercase letters and vectors are denoted by bold lowercase letters.
$\mathbb{C}^{m\times n}$ and $\mathbb{C}^{k}$ are the sets of ${m\times n}$ matrices and $k\times1$ vectors with complex-valued entries.
$\bI$ and $\bzero$  denote the identity matrix and the matrix of zeros, with the size determined by a subscript or from the context. 
Superscripts $(\cdot)^{\top}$, $(\cdot)^*$, and $(\cdot)^{\HH}$ denote the transpose, the complex conjugate, and the conjugate transpose.
The symbol $\textrm{tr}(\cdot)$ indicates the trace of a square matrix.
$\|\cdot\|_2$ denotes the Euclidean norm (for a vector) or the spectral norm (for a matrix). 
$\otimes$ represents the Kronecker product.
${\vec}(\bX)$ indicates the vector obtained by column-wise stacking of the entries of $\bX$.
$\Re(\bX)$ denotes the real part of the matrix $\bX$.
$\arg(x)$ represents the argument of $x$.
$\lambda_{\max}(\bX)$ denotes the largest eigenvalue of $\bX$, and $\be_{\max}(\bX)$ is the associated eigenvector.
$\bx\sim \mathcal{CN}(\bm,\bR)$ means that $\bx$ obeys a circularly symmetric complex Gaussian distribution with mean $\bm$ and covariance matrix $\bR$.
Finally, $\mathbb{E}(x)$ denotes the expectation of the random variable $x$.

\subsection{Acronyms}
 A list of acronyms used in this paper can be found in Table \ref{Table:1}.

\begin{table}[!htbp]
\centering
\caption{List of acronyms}
\begin{tabular}{|c|c|}
\hline
  \textbf{Acronym} & \textbf{Meaning} \\
  \hline
  RF & radio frequency \\
  MFRF & multifunction radio frequency  \\
  EW & electronic warfare\\
  AM & amplitude modulation\\
  FMCW & frequency modulated continuous wave\\
  OFDM &orthogonal frequency division multiplexing\\
  MIMO & multiple-input-multiple-output \\
  NP &  Neyman-Pearson\\
  SNR &  signal-to-noise ratio \\
   SINR & signal-to-interference-plus-noise ratio\\
   PAPR & peak-to-average-power ratio\\
   ECM & electronic countermeasure \\
   ADMM & alternating direction method of multipliers\\
   MM & majorization minimization\\
   SER &  symbol error rate\\
  \hline
\end{tabular}
\label{Table:1}
\end{table}

\section{Signal Model and Problem Formulation} \label{sec:SigModel}
\begin{figure}[!htbp]
\centerline{\includegraphics[width=0.75\textwidth]{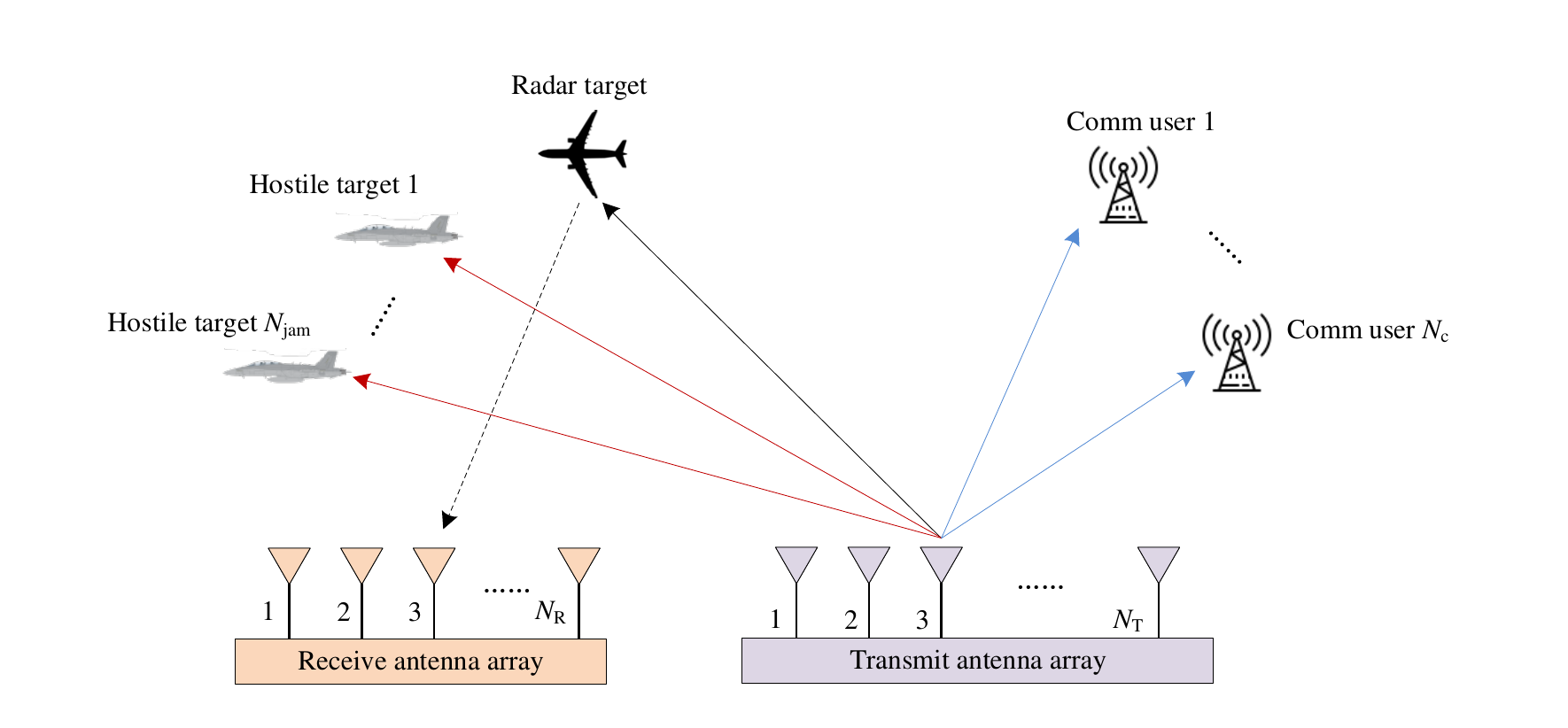}}
\caption{Illustration of a MIMO multifunction RF system.}
\label{Fig:1}
\end{figure}
Consider a MIMO system with $\NT$ transmit antennas and $\NR$ receive antennas. Let $\bS =[\bs_1, \bs_2, \cdots, \bs_{\NT}]^{\top}\in \mathbb{C}^{\NT \times L}$ denote the discrete-time baseband waveform matrix and let $\ba(\theta)$ be the transmit array steering vector for direction $\theta$, where $\bs_m$ denotes the waveform of the $m$th transmitter, and $L$ is the code length. Following \cite{Li2007mimoIntroduction}, the emitted signal at $\theta$ is given by
\begin{equation}\label{eq:emittedSig}
  \ba^\HH(\theta)\bS = \sum_{m=1}^{\NT} a_m^*(\theta) \bs_m^{\top},
\end{equation}
where $a_m^*(\theta)$ is the $m$th element of $\ba^\HH(\theta)$. The above expression shows that the emitted signal is a linear combination of the transmit waveforms, the coefficients of which vary with the angle. Therefore, utilizing the waveform diversity of MIMO systems, it is possible to transmit different signals in given directions. 

Typically, a MIMO system can be  configured to operate in various modes. In a searching mode, the MIMO system can transmit quasi-orthogonal waveforms to achieve an omnidirectional coverage. Then the system can detect the targets from the returns and estimate their parameters (including range and directions) \cite{Xu2008MIMO,Roberts2010IAA}. In a tracking mode, the MIMO system can transmit partially correlated waveforms and refine the target estimates from the echoes (see, e.g., \cite{Stoica2007Probing}).  In addition to target searching and tracking, we also assume that the system is able to identify the friendly targets using coded ``interrogating" signals.

For the scenario of interest in this paper, we assume that the MIMO system must communicate with the friendly targets in the presence of hostile targets (e.g., eavesdroppers). Meanwhile, the system must also detect a target of interest and estimate its parameters. As a result, we aim to design waveforms for this MIMO system such that it can simultaneously detect the target,  communicate with the friendly targets (we call them communication users in the sequel), and jam the hostile targets, as illustrated in Fig. \ref{Fig:1}.
Assume that $N_\textrm{c}$ communication receivers and $N_\textrm{jam}$  hostile targets are present.
Let $\bTheta_\textrm{c} =\{\theta_{\textrm{c},1},\theta_{\textrm{c},2},\cdots,\theta_{\textrm{c},N_\textrm{c}}\}$ denote the set of the directions of the communication receivers. Then the signals transmitted to the communication receivers can be written as
\begin{equation}
  \bY_\textrm{c} = \bA^\HH(\bTheta_\textrm{c}) \bS,
\end{equation}
where $\bA(\bTheta_\textrm{c}) = [\ba(\theta_{\textrm{c},1}), \ba(\theta_{\textrm{c},2}),\cdots, \ba(\theta_{\textrm{c},N_\textrm{c}})] \in \mathbb{C}^{\NT \times N_\textrm{c}}$, $\bY_\textrm{c}=[\by_{\textrm{c},1},\by_{\textrm{c},2},\cdots,\by_{\textrm{c},N_\textrm{c}}]^\top$, and $\by_{\textrm{c},n}$ represents the communication signal sent to the $n$th receiver, $n=1,2, \cdots, N_\textrm{c}$. To communicate with these receivers, we aim to design $\bS$ such that
\begin{align}\label{eq:CommMatch}
\bA^\HH(\bTheta_\textrm{c}) \bS \approx \bD_\textrm{c},
\end{align}
where $\bD_\textrm{c} = [\bd_{\textrm{c},1},\bd_{\textrm{c},2},\cdots, \bd_{\textrm{c},N_\textrm{c}}]^{\top}\in \mathbb{C}^{N_\textrm{c}\times L}$, and $\bd_{\textrm{c},n}$ denotes the desired signal for the $n$th communication receiver, $n=1,2, \cdots, N_\textrm{c}$.

Similarly, the signals transmitted toward the hostile targets can be written as
\begin{equation}
  \bA^\HH(\bTheta_\textrm{jam}) \bS,
\end{equation}
where $\bTheta_\textrm{jam} = \{\theta_{\textrm{jam},1}, \theta_{\textrm{jam},2},\cdots, \theta_{\textrm{jam},N_\textrm{jam}}\} $ is the set of the directions of the hostile targets, and $\bA(\bTheta_\textrm{jam}) \in \mathbb{C}^{\NT \times N_\textrm{jam}}$ is defined similarly to $\bA(\bTheta_\textrm{c})$. Thus, to jam the hostile targets, $\bS$ should satisfy
\begin{align}\label{eq:JamMatch}
\bA^\HH(\bTheta_\textrm{jam}) \bS \approx \bD_\textrm{jam},
\end{align}
where $\bD_\textrm{jam} = [\bd_{\textrm{jam},1},\bd_{\textrm{jam},2},\cdots, \bd_{\textrm{jam},N_\textrm{jam}}]^\top \in \mathbb{C}^{N_\textrm{jam}\times L}$, and $\bd_{\textrm{jam},m}$ denotes the signal intended for the $m$th hostile target, $m=1,2, \cdots, N_\textrm{jam}$.

Let $\theta_\textrm{t}$ denote the target direction. Then the target signals received by the MIMO systems can be written as
\begin{align}\label{eq:yt}
  \bY_\textrm{t} = \alpha_\textrm{t} \bb(\theta_\textrm{t}) \ba^\HH(\theta_\textrm{t})\bS + \bN,
\end{align}
where $\bb(\theta_\textrm{t}) \in \mathbb{C}^{\NR \times 1}$ is the receive array steering vector for $\theta_\textrm{t}$, $\bN\in \mathbb{C}^{\NR \times L}$ is the disturbance matrix, and $\alpha_\textrm{t}$ denotes the complex target amplitude. We assume for simplicity that the clutter can be ignored such that the disturbance is signal-independent. Such an assumption is justified if  the energy of the signals transmitted toward the communication users and the hostile targets is small (compared with that of the signals transmitted toward the target). More importantly, this assumption is also justified by the analysis following equation \eqref{eq:SINR} below.

Letting $\by_\textrm{t} =\vec(\bY_\textrm{t})$ and using the fact that $\vec(\bA\bB\bC) = (\bC^\top \otimes \bA) \vec(\bB)$ \cite{HornJohnson1990matrixbook}, we can write \eqref{eq:yt} as
\begin{align}\label{eq:TargetModel}
  \by_\textrm{t} = \alpha_\textrm{t} \bH(\theta_\textrm{t}) \bs + \bn,
\end{align}
where $\bH(\theta_\textrm{t}) = \bI_L \otimes \bb(\theta_\textrm{t}) \ba^\HH(\theta_\textrm{t})$, $\bs = \vec(\bS)$, and $\bn = \vec(\bN)$.
We can now formulate the following binary hypothesis testing problem for target detection:
\begin{equation}
  \begin{cases}
    \mathcal{H}_0:& \by=  \bn, \\
    \mathcal{H}_1:& \by= \alpha_\textrm{t} \bH(\theta_\textrm{t}) \bs +\bn.
  \end{cases}
\end{equation}
Assume that $\bn \sim \mathcal{CN}(\bzero,\bR)$. According to the Neyman-Pearson criterion \cite{kaybook1998}, we decide $\mathcal{H}_1$ if
\begin{equation}
  \Re(\alpha_\textrm{t}\by^\HH \bH(\theta_\textrm{t}) \bs ) > T_h,
\end{equation}
where $T_h$ is the detection threshold. The detection probability of this detector is given by \cite{kaybook1998}
\begin{equation} \label{eq:Pd}
  P_\textrm{D} = \frac{1}{2} \textrm{erfc}\left\{\textrm{erfc}^{-1}(2P_{\textrm{FA}}) - \sqrt{\textrm{SINR}}) \right\},
\end{equation}
 where $\textrm{erfc}(x)=\frac{2}{\sqrt{\pi}}\int_{x}^{\infty}e^{-t^2}\textrm{d}t$ is the complementary error function, $P_{\textrm{FA}}$ is the probability of false alarm,
and the SINR is defined by
\begin{align}\label{eq:SINR}
  \textrm{SINR} = |\alpha_\textrm{t}|^2 \bs^\HH\bH^\HH(\theta_\textrm{t}) \bR^{-1} \bH(\theta_\textrm{t}) \bs.
\end{align}
We note that the definition \eqref{eq:SINR} coincides with the SINR at the output of an optimum spatial-temporal filter. To see this, let $\bw \in \mathbb{C}^{LN_\txR \times 1} $ denote the spatial-temporal filter at the receiver. According to the signal model in \eqref{eq:TargetModel}, the SINR that corresponds to $\bw$ is given by
\begin{align*}
  \textrm{SINR}
  &= \frac{|\alpha_\textrm{t} \bw^\dagger \bH(\theta_\textrm{t}) \bs|^2}{\bw^\dagger\mathbb{E}(\bn\bn^\dagger)\bw} =\frac{|\alpha_\textrm{t} \bw^\dagger \bH(\theta_\textrm{t}) \bs|^2}{\bw^\dagger\bR\bw}.
\end{align*}
It can be checked that
\begin{equation*}
  \bw^{\textrm{opt}} = \alpha_0 \bR^{-1} \bH(\theta_\textrm{t}) \bs
\end{equation*}
maximizes the SINR ($\alpha_0$ is an arbitrary nonzero constant), and the associated maximum SINR is given by \eqref{eq:SINR}. This fact implies that \eqref{eq:SINR} is rather insensitive to  clutter. Indeed, the filter $\bw^{\textrm{opt}} $ will adaptively form deep nulls at the directions of the clutter and consequently, we will reduce the clutter power to a low level.

Note from \eqref{eq:Pd} that  $P_\textrm{D}$ is a monotonically increasing function of $\textrm{SINR}$. Therefore, the maximization of  $\textrm{SINR}$ is equivalent to the maximization of the detection probability. Making use of this observation as well as the constraints in \eqref{eq:CommMatch} and \eqref{eq:JamMatch}, we formulate the following optimization problem to maximize the target detection performance of the MIMO multifunction system:
\begin{align} \label{eq:DesignEnergy}
  \max_{\bs} & \ \bs^\HH \bM \bs \nonumber \\
  \textrm{s.t.} & \ \bA^\HH(\bTheta) \bS = \bD,  \bs^\HH\bs = e_t,
\end{align}
where $\bM = \bH^\HH(\theta_\textrm{t}) \bR^{-1} \bH(\theta_\textrm{t})$, $\bA(\bTheta) =[\ba(\theta_1),\ba(\theta_2),\cdots,\ba(\theta_{N_0})] =[\bA(\bTheta_\textrm{c}), \bA(\bTheta_\textrm{jam})] \in \mathbb{C}^{\NT \times N_0}$, $\bD = [\bd_1, \bd_2, \cdots, \bd_{N_0}]^\top = [\bD_\textrm{c}^\top, \bD_\textrm{jam}^\top]^\top \in \mathbb{C}^{N_0\times L}$, $N_0 = N_\textrm{c}+N_\textrm{jam}$, and $e_t$ is the available transmit energy. Moreover, to make the optimization problem in \eqref{eq:DesignEnergy} feasible, we assume in the following that $N_\txT>N_0$.

\emph{Remark 1:} In the formulation of  \eqref{eq:DesignEnergy}, we have assumed that $\bR$ and $\bTheta$ are known. $\bR$ can be estimated from target-free secondary data. Regarding $\bTheta$, we assume that the system has access to estimates of the locations of the communication receivers and hostile targets obtained in the previous stages. To update the information about $\bTheta$, the system can transmit probing waveforms and estimate $\bTheta$ from the echoes (e.g., by the methods in \cite{Stoica2007Probing, Xu2008MIMO,Roberts2010IAA}).

\emph{Remark 2:} For a radar-only system based on the same array, the achievable SINR can be obtained by ignoring the multifunction constraint (i.e., the first equality constraint in \eqref{eq:DesignEnergy}) and  solving the following problem:
\begin{align} \label{eq:RadarOnly}
  \max_{\bs} & \ \bs^\HH \bM \bs \nonumber \\
  \textrm{s.t.} & \ \bs^\HH\bs = e_t.
\end{align}
Considering \eqref{eq:RadarOnly}, it is evident that the maximum SINR of a radar-only system is $e_t \lambda_{\max}(\bM)$, which is achieved for $\bs = \sqrt{e_t}\be_{\max}(\bM)$.

\emph{Remark 3:} In addition to the SINR performance of the MFRF system, one may also be interested in the information rate and the electronic countermeasure (ECM) capability that can be achieved by the system. According to \cite{Larsson2013precoding}, for a  Gaussian broadcast channel, if the input information bits are drawn from a Gaussian alphabet, the achievable sum-rate for the communication users can be written as
\begin{equation}\label{eq:sumrate}
  R = \sum_{n=1}^{N_\textrm{c}} \log_2 (1+\chi_n),
\end{equation}
where $\chi_n$ is the SINR associated with the $n$th communication receiver:
\begin{equation}\label{eq:CommSINR}
  \chi_n = \frac{\mathbb{E}|d_{\textrm{c},n}(k)|^2}{\mathbb{E}|\ba^\HH(\theta_{\textrm{c},n})\bS(:,k)-d_{\textrm{c},n}(k)|^2 + \sigma_n^2}.
\end{equation}
Here $d_{\textrm{c},n}(k)$ is the $k$th element of $\bd_{\textrm{c},n}$, $\bS(:,k)$ denotes the $k$th column of $\bS$, $\mathbb{E}|\ba^\HH(\theta_{\textrm{c},n})\bS(:,k)-d_{\textrm{c},n}(k)|^2$ stands for the multiuser interference, and $\sigma_n^2$ is the noise power for the $n$th communication receiver. Note that if the constraint in \eqref{eq:DesignEnergy} is satisfied, i.e., the synthesized signals perfectly match the desired signals, $\chi_n$ reduces to
\begin{equation}
  \chi_n = \frac{\mathbb{E}|d_{\textrm{c},n}(k)|^2}{ \sigma_n^2} \triangleq \textrm{CSNR}.
\end{equation}
Thus, the achievable sum-rate is determined by the signal-to-noise ratio (SNR) of the desired communication signals (i.e., CSNR).
Regarding the ECM capability, it is usually determined by the  energy of the  jamming signals, i.e., the energy transmitted toward the hostile targets. Hence, we can use $\|\ba^\HH(\theta_{\textrm{jam},m}) \bS\|_2^2$ (i.e., $\|\bd_{\textrm{jam},m}\|_2^2$, if the constraint in \eqref{eq:DesignEnergy} is satisfied) to measure the ECM capability of the MFRF system, $m=1,2, \cdots, N_\textrm{jam}$.
\section{Optimal Waveforms For General $\bR$} \label{sec:OptimalSolution}

It is a well-known result that any matrix $\bS$ satisfying the constraint $\bA^\HH(\bTheta) \bS = \bD$ can be written as \cite[Result R29]{stoica2005spectral}
\begin{equation}\label{eq:S}
  \bS = \hat{\bS} + \bB \bV,
\end{equation}
where $\hat{\bS} = \bA(\bTheta)(\bA^\HH(\bTheta)\bA(\bTheta))^{-1}\bD$, $\bB \in \mathbb{C}^{\NT \times (\NT-N_0)}$ is a unitary matrix that spans the null space of $\bA^\HH(\bTheta)$ (i.e., $\bA^\HH(\bTheta)\bB = \bzero$ and $\bB^\HH\bB = \bI$), and $\bV \in \mathbb{C}^{ (\NT-N_0)\times L}$ is arbitrary (note that, for $\NT>N_0$, the inverse matrix in $\hat{\bS}$ exists under mild conditions).
As a consequence, making using of \eqref{eq:S}, we can eliminate the first equality constraint in \eqref{eq:DesignEnergy}.
Also, the transmit energy can be rewritten as
\begin{align}
  \bs^\HH \bs &= \tr(\bS^\HH\bS) \nonumber \\
                        &= \tr(\hat{\bS}^\HH\hat{\bS}) + \tr(\bV^\HH \bV)\nonumber \\
                        &= \hat{\bs}^\HH \hat{\bs} + \bv^\HH \bv,
\end{align}
where $\hat{\bs} = \vec(\hat{\bS})$, and $\bv = \vec(\bV)$.
Thus, the maximization problem in \eqref{eq:DesignEnergy} can be reformulated as
\begin{align} \label{eq:DesignEnergy2}
  \max_{\bv} & \ (\hat{\bs} + \hat{\bB} \bv )^\HH \bM (\hat{\bs} + \hat{\bB} \bv ) \nonumber \\
  \textrm{s.t.} & \ \bv^\HH\bv = \hat{e}_t,
\end{align}
where  $\hat{\bB} = \bI_L \otimes \bB$, and $\hat{e}_t = e_t - \hat{\bs}^\HH \hat{\bs}$. Note that $\hat{e}_t$ must be a positive number. Hence, to ensure that  the waveform design problem is feasible, the transmit energy $e_t$ should be larger than the energy of $\hat{\bs}$, which is $\tr(\bD^\HH(\bA^\HH(\bTheta)\bA(\bTheta))^{-1}\bD)$.
It follows, for example, from \cite{Huang2007decomposition} that the optimization problem in \eqref{eq:DesignEnergy2} is hidden-convex, and thus it can be solved by the Lagrange multiplier method. The Lagrangian for \eqref{eq:DesignEnergy2} is the following:
\begin{equation}
  L_\nu(\bv) = -(\hat{\bs} + \hat{\bB} \bv )^\HH \bM (\hat{\bs} + \hat{\bB} \bv ) + \nu(\bv^\HH\bv - \hat{e}_t),
\end{equation}
where $\nu$ is the multiplier associated with the energy constraint. Consequently, the optimal solution, obtained by setting the first-order derivative of $ L_\nu(\bv)$ w.r.t. $\bv$ to zero, is given by
\begin{equation}
  \bv^\star = (\nu^\star \bI - \hat{\bB}^\HH\bM\hat{\bB})^{-1}\hat{\bB}^\HH\bM\hat{\bs},
\end{equation}
where $\nu^\star > \lambda_{\max}(\hat{\bB}^\HH\bM\hat{\bB})$  is obtained by solving the following equation: 
\begin{equation}\label{eq:solvenu}
\hat{\bs}^\HH\bM \hat{\bB}(\nu \bI - \hat{\bB}^\HH\bM\hat{\bB})^{-2}\hat{\bB}^\HH\bM\hat{\bs} = \hat{e}_t.
\end{equation}

To solve \eqref{eq:solvenu}, we let $\hat{\bB}^\HH\bM\hat{\bB} = \bU \boldsymbol{\Sigma} \bU^\HH$ be the eigenvalue decomposition of $\hat{\bB}^\HH\bM\hat{\bB} $, where $\bU \in \mathbb{C}^{M \times M}$ is a unitary matrix, $\boldsymbol{\Sigma} \in \mathbb{C}^{M \times M}$ is a diagonal matrix,  and $M=(\NT-N_0)L$.

Thus, letting $\tilde{\bs} = \bU^\HH\hat{\bB}^\HH\bM\hat{\bs}$, we can rewrite \eqref{eq:solvenu} as
\begin{equation}\label{eq:solvenu2}
  \sum_{m=1}^{M } \frac{|\tilde{s}(m)|^2}{(\nu - \tau_m)^2}= \hat{e}_t,
\end{equation}
where $\tilde{s}(m)$ is the $m$th element of $\tilde{\bs}$, and $\tau_m$ is the $m$th diagonal element of $\boldsymbol{\Sigma}$ (which we assume without loss of generality that $\tau_1\geq \tau_2 \geq \cdots \geq \tau_{M}$). Recall that $\nu^\star > \lambda_{\max}(\hat{\bB}^\HH\bM\hat{\bB}) = \tau_1$, and note that the left hand side of \eqref{eq:solvenu2}, denoted by $f(\nu)$, is a monotonically decreasing function for $\nu \in (\tau_1, \infty)$. Moreover, as $\nu \rightarrow \tau_1$, $f(\nu)\rightarrow \infty$, and as $\nu \rightarrow \infty$, $f(\nu)\rightarrow 0$. We can obtain that there is a unique solution for $\nu \in (\tau_1, \infty)$. To find the solution of interest, we also note that
\[ \frac{\|\tilde{\bs}\|_2^2}{(\nu^\star - \tau_M)^2} \leq \hat{e}_t \leq \frac{\|\tilde{\bs}\|_2^2}{(\nu^\star - \tau_1)^2},\]
indicating that
\begin{equation}
  \frac{\|\tilde{\bs}\|_2}{\sqrt{\hat{e}_t}}+{\tau_M} \leq \nu^\star \leq \frac{\|\tilde{\bs}\|_2}{\sqrt{\hat{e}_t}}+{\tau_1}.
\end{equation}
Using the above observation, we can use, e.g., a bisection method or a Newton's method, to easily solve \eqref{eq:solvenu}.

\section{Optimal Waveforms For Structured $\bR $} \label{sec:structured}
To gain more insights into the optimal solution to \eqref{eq:DesignEnergy}, we assume that the columns of $\bN$ are independent and identically distributed and have a covariance matrix denoted $\bar{\bR}$ (similar to assumptions made in \cite{Xu2008MIMO,Liu2015NoTraining,Liu2018Tunable}). Hence, the covariance matrix of $\bn = \vec(\bN)$ can be written as $\bR = \bI_L \otimes \bar{\bR}$. Note that such an assumption is justified, e.g.,  when the disturbance is dominated by internal white noise and external hostile jamming signals. In this situation, $\bN$ can be modeled as
\begin{align}\label{eq:disturbanceModel}
  \bN = \bN_\textrm{I} +\sum_{n=1}^{N_J} \bb(\theta_{\textrm{j},n}) \bs^\top_{\textrm{j},n},
\end{align}
where $\bN_\textrm{I}$ denotes the internal white noise, $N_J$ is the number of jammers, and $\theta_{\textrm{j},n}$ and $\bs_{\textrm{j},n}$ denote the direction and the jamming signals of the $n$th jammer, $n=1,\cdots,N_J$. Then
\begin{align}\label{eq:disturbanceModel2}
  \bn = \vec(\bN) = \bn_\textrm{I} +\sum_{n=1}^{N_J} \bs_{\textrm{j},n}\otimes\bb(\theta_{\textrm{j},n}),
\end{align}
where $\bn_\textrm{I} = \vec(\bN_\textrm{I})$. Assuming that the jamming signals are white and uncorrelated with one another, we have
\begin{align}
  \bR = \mathbb{E}(\bn\bn^\HH)
  &= \sigma^2 \bI_{L\NR} + \sum_{n=1}^{N_J} \sigma_{\textrm{j},n}^2 \bI_L \otimes (\bb(\theta_{\textrm{j},n}) \bb^\HH(\theta_{\textrm{j},n})) \nonumber \\
  &= \bI_L \otimes ( \sigma^2 \bI_{\NR} + \sum_{n=1}^{N_J} \sigma_{\textrm{j},n}^2 \bb(\theta_{\textrm{j},n}) \bb^\HH(\theta_{\textrm{j},n})) \nonumber \\
  &\triangleq \bI_L \otimes \bar{\bR},
\end{align}
where $\sigma^2$ is the power of the white noise, and $\sigma_{\textrm{j},n}^2$ is the power of the $n$th jammer, $n=1,\cdots,N_J$.
Using the above expression for $\bR$, we can write $\bM$ as
\begin{align}
  \bM
  &= \bH^\HH(\theta_\textrm{t}) \bR^{-1} \bH(\theta_\textrm{t}) \nonumber \\
  &= \bI_L \otimes (\bb^\HH(\theta_\textrm{t})\bar{\bR}^{-1}\bb(\theta_\textrm{t}) \ba(\theta_\textrm{t})\ba^\HH(\theta_\textrm{t}) ) ,
\end{align}
By using the identity $\tr(\bA\bB\bC\bD) = \vec^\top(\bD)(\bA \otimes \bC^\top) \vec(\bB^\top)$ \cite{bernstein2009matrix}, we obtain
\begin{align}
  \textrm{SINR} &= \bs^\HH \bM \bs \nonumber \\
  &=\ba^\HH(\theta_\textrm{t})\bS \bS^\HH \ba(\theta_\textrm{t}) \cdot \bb^\HH(\theta_\textrm{t})\bar{\bR}^{-1}\bb(\theta_\textrm{t}) \nonumber \\
  &= \textrm{SINR}_{\txT} \cdot  \textrm{SINR}_{\txR},
\end{align}
where we define $\textrm{SINR}_{\txT} = \ba^\HH(\theta_\textrm{t})\bS \bS^\HH \ba(\theta_\textrm{t})$ as the transmit SINR (also called the transmit beampattern at $\theta_\textrm{t}$), and $\textrm{SINR}_{\txR} = \bb^\HH(\theta_\textrm{t})\bar{\bR}^{-1}\bb(\theta_\textrm{t})$ as the receive SINR.
 Therefore, in the present case, the waveform design problem can be reformulated as
\begin{align} \label{eq:DesignEnergySimple}
  \max_{\bs} & \ \ba^\HH(\theta_\textrm{t})\bS \bS^\HH \ba(\theta_\textrm{t}) \nonumber \\
  \textrm{s.t.} & \ \bA^\HH(\bTheta) \bS = \bD,  \bs^\HH\bs = e_t.
\end{align}
Using \eqref{eq:S}, we have $\bS^\HH \ba(\theta_\textrm{t}) = \bq + \bV^\HH \bu$, where $\bq = \hat{\bS}^\HH \ba(\theta_\textrm{t})$, and $\bu = \bB^\HH\ba(\theta_\textrm{t})$. Then the problem in \eqref{eq:DesignEnergySimple} becomes
\begin{align} \label{eq:DesignEnergySimple2}
\max_{\bV} & \  (\bq + \bV^\HH \bu)^\HH (\bq + \bV^\HH \bu) \nonumber \\
  \textrm{s.t.}& \  \tr(\bV\bV^\HH) = \hat{e}_t.
\end{align}
The objective in \eqref{eq:DesignEnergySimple2} can be rewritten as
\begin{equation}
 \bu^\HH\bV\bV^\HH\bu + 2\Re(\bu^\HH\bV\bq) + \bq^\HH\bq.
\end{equation}
Note that
\begin{align}
  \bu^\HH\bV\bV^\HH\bu
  = \|\bV^\HH\bu\|_2^2 
  \leq  \| \bV \|_2^2 \|\bu\|_2^2
  \leq \hat{e}_t \bu^\HH \bu.
\end{align}
The upper bound is achieved for
\begin{equation}\label{eq:optV1}
  \bV = \sqrt{\hat{e}_t} \bar{\bu} \bx^\HH,
\end{equation}
where $\bar{\bu} = \bu/ \|\bu\|_2$, and $\bx \in \mathbb{C}^{L \times 1}$ is an arbitrary normalized vector (i.e., $\bx^\HH \bx =1$). In addition, we can verify that 
\begin{align} \label{eq:proof1}
  \Re(\bu^\HH\bV\bq)
  \leq \|\bu\| _2\| \bV \|_2\|\bq\|_2
\leq \sqrt{\hat{e}_t}\|\bu\| _2\|\bq\|_2,
\end{align}
and the equality holds for
\begin{equation}\label{eq:optV2}
  \bV = \sqrt{\hat{e}_t} \bar{\bu} \bar{\bq}^\HH,
\end{equation}
where $\bar{\bq} = \bq /\|\bq\|_2$.
Combining the results in \eqref{eq:optV1} and \eqref{eq:optV2}, we conclude that the maximum of \eqref{eq:DesignEnergySimple2} is achieved for the $\bV$ given by \eqref{eq:optV2}. Therefore, the optimal waveform matrix can be written as
\begin{align}\label{eq:WaveformUnderEnergyConstraint}
  \bS
  &= \hat{\bS} + \sqrt{\hat{e}_t} \bB \bar{\bu} \bar{\bq}^\HH \nonumber \\
  &= \hat{\bS} + \beta_0 \bB\bB^\HH \ba(\theta_\textrm{t})\ba^\HH (\theta_\textrm{t})\hat{\bS} \nonumber \\
  &=(\bI +\beta_0 \bB\bB^\HH \ba(\theta_\textrm{t})\ba^\HH (\theta_\textrm{t}) )\hat{\bS},
\end{align}
where $\beta_0 = \sqrt{\hat{e}_t}/(\|\bu\|_2\|\bq\|_2)$. The associated objective value is given by
\begin{equation}\label{eq:SINRT}
  \textrm{SINR}_\txT = \ba^\HH(\theta_\textrm{t})\hat{\bS} \hat{\bS}^\HH \ba(\theta_\textrm{t}) (1+\beta_0 \ba^\HH(\theta_\textrm{t})\bB \bB^\HH\ba(\theta_\textrm{t}))^2.
\end{equation}
Note that $\ba^\HH(\theta_\textrm{t})\hat{\bS} \hat{\bS}^\HH \ba(\theta_\textrm{t}) = \|\bq\|_2^2$ and $\ba^\HH(\theta_\textrm{t})\bB \bB^\HH\ba(\theta_\textrm{t}) = \|\bu\|_2^2$. Thus, $\textrm{SINR}_\txT$ can also be written as
\begin{equation}\label{eq:SINRT2}
\textrm{SINR}_\txT = (\|\bq\|_2 + \sqrt{\hat{e}_t}\|\bu\|_2)^2.
\end{equation}
\subsection{The Rank of Optimal Waveform Matrix}
From \eqref{eq:WaveformUnderEnergyConstraint}, we can see that the rank of $\bS$ satisfies
  \begin{align}
    \textrm{rank}(\bS) \leq \textrm{rank}(\hat{\bS}) \leq \textrm{rank}(\bD) \leq N_0.
  \end{align}
  As a result, the optimal waveform matrix tends to have a low rank (under the working assumption that $N_0 < \NT$). In particular, for the case of $N_0 =1$ (which corresponds to a dual-function system integrating radar detection and either communication or jamming), $\textrm{rank}(\bS) = 1$, indicating that the dual-function system should transmit coherent waveforms to maximize the SINR (In other words, the phased array maximizes the SINR; see \cite{li2010phased} for other cases in which the phased array maximizes the SNR.). 
  In this case,  $\bS$ can be written as (see the Appendix for a derivation)
  \begin{equation}\label{eq:CoherentWaveform}
    \bS = \bw \bd^\top,
  \end{equation}
  where $\bw = \alpha_1\ba(\bar{\theta})  + \alpha_2 \ba(\theta_\textrm{t})$, $\bar{\theta}$ denotes the direction of the communication receiver or the hostile target, $\ba(\bar{\theta}) \triangleq \bA(\bTheta)$, $\bd \triangleq \bD^\top$ (note that $\bA(\bTheta)$ and $\bD^\top$ are vectors in this case), $\alpha_1 = N_\textrm{T}^{-1} - \beta_0 G^2(\theta_\textrm{t}, \bar{\theta})$, $\alpha_2=\beta_0 B^*(\theta_\textrm{t}, \bar{\theta})$, $B(\theta_\textrm{t}, \bar{\theta}) = {\ba^\HH(\bar{\theta})\ba(\theta_\textrm{t})}/{\NT}$ is the normalized beampattern of the transmit array at $\bar{\theta}$ when the array is pointing to $\theta_\textrm{t}$, and $G(\theta_\textrm{t}, \bar{\theta}) = |B(\theta_\textrm{t}, \bar{\theta})|$ is the associated normalized gain. Observe from \eqref{eq:CoherentWaveform} that each of the coherent waveforms is a scaled version of the desired waveform $\bd$, and the associated beamformer $\bw$ is a linear combination of $\ba(\bar{\theta})$ and $\ba(\theta_\textrm{t})$.
\subsection{The Achievable SINR} \label{Subsec:achievableSINR}
Using \eqref{eq:SINRT}, we can show after straightforward calculations that
\begin{equation}
   \ba^\HH(\theta_\textrm{t})\hat{\bS} \hat{\bS}^\HH \ba(\theta_\textrm{t}) \leq \textrm{SINR}_\txT \leq e_t N_\txT,
\end{equation}
where the upper bound (i.e., $e_t N_\txT$) is the SINR achieved by a radar-only system (that is, a radar system working in a phased-array mode).
Moreover, if we assume that the transmit array is sufficiently large (i.e., $\NT \gg 1$) and the angles $\{\theta_k\}_{k=1}^{N_0}$ are not too close to one another, then \cite{Stoica1989CRB}
\begin{equation}
  \bA^\HH(\bTheta)\bA(\bTheta) \approx \NT \bI.
\end{equation}
As a result,
\begin{align}
  \bB\bB^\HH
  & = \bI - \bA(\bTheta)(\bA^\HH(\bTheta)\bA(\bTheta))^{-1}\bA^\HH(\bTheta) \nonumber \\
  & \approx \bI - N_\textrm{T}^{-1} \bA(\bTheta)\bA^\HH(\bTheta)
\end{align}
and
\begin{align}
  \|\bu\|_2^2 = \ba^\HH(\theta_\textrm{t})\bB \bB^\HH\ba(\theta_\textrm{t}) \approx \NT(1-G_{\textrm{sos}} ),
\end{align}
where $G_{\textrm{sos}} = \sum\nolimits_{k=1}^{N_0} G_k^2$, and $G_k = G(\theta_\textrm{t}, {\theta_k})$ is the normalized gain of the array at $\theta_k$ when pointing to $\theta_\textrm{t}$ \footnote{Note that equation (46) implicitly assumes that $G_{\textrm{sos}} \leq 1$, implying that the approximations made in this section might not hold if two of the communication receivers/hostile targets fall within the mainlobe. Also, here sos stands for the sum of squares.}. Moreover,
\begin{align}
  \|\bq\|_2^2 \approx N_\textrm{T}^{-2} \ba^\HH(\theta_\textrm{t})\bA(\bTheta)\bD\bD^\HH\bA^\HH(\bTheta) \ba(\theta_\textrm{t}).
\end{align}
Assume that the desired waveforms are such that $\bd_m^\dagger \bd_n \approx 0, m\neq n, 1\leq m,n \leq N_0$. Then $\bD\bD^\HH \approx\textrm{Diag}([\bar{e}_1,\bar{e}_2, \cdots,\bar{e}_{N_0}]^\top)$, where $\bar{e}_k = \|\bd_k\|_2^2$ is the energy of the $k$th desired waveform, $k=1,2,\cdots,N_0$. Under this assumption,
\begin{align}
  \|\bq\|_2^2 \approx \sum_{k=1}^{N_0} \bar{e}_k G_k^2.
\end{align}
Therefore, $\textrm{SINR}_\txT$ can be approximated by (see \eqref{eq:SINRT2}):
\begin{equation}\label{eq:SINRApprox}
  \textrm{SINR}_\txT \approx \left(\sqrt{\sum\nolimits_{k=1}^{N_0} \bar{e}_k G_k^2}+ \sqrt{\hat{e}_t\NT(1-G_{\textrm{sos}} )}\right)^2.
\end{equation}
$\bullet$ Case 1: The communication receivers and the hostile targets all fall on sidelobes such that the normalized gain $G_k\approx 0, k=1, 2,\cdots, N_0$. In this case, $\textrm{SINR}_\txT$ can be simplified to
\begin{equation}
  \textrm{SINR}_\txT \approx \hat{e}_t\NT.
\end{equation}
Note that
\begin{align}
  \hat{e}_t
  &=e_t - \hat{\bs}^\HH \hat{\bs} \approx e_t - N_\txT^{-1} \sum_{k=1}^{N_0} \bar{e}_k.
\end{align}
Hence, if the sum of the energy of the desired signals is large, the achievable SINR of the MFRF system can be much smaller than that of the radar-only system.

$\bullet$ Case 2: $\bar{e}_1=\bar{e}_2=\cdots=\bar{e}_{N_0}\triangleq \bar{e}$. In this case, $\textrm{SINR}_\txT$ can be approximated by
\begin{equation}\label{eq:equalEnergy}
  \textrm{SINR}_\txT \approx \left(\sqrt{\bar{e}G_{\textrm{sos}}} + \sqrt{\hat{e}_t\NT (1-G_{\textrm{sos}})} \right)^2.
\end{equation}
Let $G_{\textrm{sos}} = \sin^2 \alpha$. Then $\textrm{SINR}_\txT \approx  (\sqrt{\bar{e}} \sin\alpha + \sqrt{\hat{e}_t\NT} \cos\alpha )^2$ and it can be checked that
\begin{equation}\label{eq:SINRbound}
  \textrm{SINR}_\txT \leq e_t\NT - (N_0-1)\bar{e},
\end{equation}
where the equality holds if $G_{\textrm{sos}} = {\bar{e}}/(\NT e_t- (N_0-1)\bar{e})$. Therefore, if $N_0 \geq 2$, the multifunction system will always suffer some SINR loss. Interestingly, a dual-function system (which corresponds to $N_0 =1$) can attain the same SINR as the radar-only system (i.e., there is no SINR loss).

$\bullet$ Case 3: One of the communication receivers or hostile targets is close to the mainlobe, and the others are on sidelobes, i.e., we assume without loss of generality that $G_1 \approx 1$, and $G_k\approx 0, k=2, \cdots, N_0$. Then we have
\begin{equation}\label{eq:SINRmainlobe}
  \textrm{SINR}_\txT \approx \bar{e}_1,
\end{equation}
which indicates that the SINR loss can be significant if $\bar{e}_1$ is small.

\section{Optimal Waveforms Under PAPR Constraint} \label{sec:PAPR}
In practical systems, for the radio frequency amplifier to operate at maximum efficiency and avoid nonlinear effects in transmitters, waveforms with low PAPR are desirable. To control the PAPR of the waveforms, we enforce the constraint
\begin{equation}\label{eq:PAPRconstraint2}
       \bs_n^\HH\bs_n = e_t/\NT, \text{PAPR}(\bs_n) \leq \rho, n = 1, \cdots, \NT,
     \end{equation}
     where $1\leq \rho \leq L$,  the PAPR of $\bs_n$ is defined as follows: 
      \begin{equation}
        \text{PAPR}(\bs_n) = \frac{\max_{l}|s_n(l)|^2}{\frac{1}{L}\sum_{l=1}^{L}|s_n(l)|^2}, n = 1, \cdots, \NT,
      \end{equation}
and $s_n(l)$ is the $l$th element of $\bs_n$.
In the special case of $\rho=1$, the PAPR constraint in \eqref{eq:PAPRconstraint2} becomes the constant-modulus constraint:
     \begin{equation}\label{eq:constantModulus}
       |s_n(l)| = {a_s}, n=1,\cdots,\NT, l = 1, \cdots, L,
     \end{equation}
 where $a_s = \sqrt{e_t/(L\NT)}$.

To design optimal waveforms under the PAPR constraint for the MIMO multifunction system, we formulate the following problem:
\begin{subequations}\label{eq:SigDesign}
  \begin{align}
  \max_{\bs} & \ \bs^\HH \bM \bs  \\
  \textrm{s.t.} & \ \|\ba^\HH(\theta_{k}) \bS - \bd_{k}^\top\|_2^2 \leq \varepsilon_k, k=1,\cdots,  N_0,  \\
  &\ \bs_n^\HH\bs_n = e_t/\NT, \text{PAPR}(\bs_n) \leq \rho, n = 1, \cdots, \NT,
\end{align}
\end{subequations}
where $ \varepsilon_k$ is a user-defined upper bound on the matching error between  $\ba^\HH(\theta_{k}) \bS$ and $\bd_{k}$, $k=1,2,\cdots,N_0$.

Note that
\begin{equation} \label{eq:equality}
  \vec(\ba^\HH(\theta_k)\bS) = \bG^\HH(\theta_k)\bs,
\end{equation}
where $\bG(\theta_k) = \bI_L \otimes \ba(\theta_k)$. Therefore, the optimization problem in \eqref{eq:SigDesign} can be reformulated as
\begin{align}\label{eq:DesignIntegratedSystem2}
\max_{\bs,t} & \ t \nonumber \\
\textrm{s.t.}
& \ \|{\bG}^\HH(\theta_k)\bs  - \bd_k\|_2^2 \leq \varepsilon_k, k=1,\cdots,  N_0, \nonumber \\
&\ \bs^\HH \bM \bs \geq t, \nonumber \\
& \ \bs_n^\HH\bs_n = e_t/\NT, \text{PAPR}(\bs_n) \leq \rho, n = 1, \cdots, \NT,
\end{align}
where $t$ is an auxiliary variable.

We will use an alternating direction method of multipliers (ADMM)-based approach to tackle the optimization problem in \eqref{eq:DesignIntegratedSystem2} (we refer to \cite{Boyd2011ADMM} for a comprehensive survey on ADMM and its applications). To this end, we define $\bM_r =\bM^{1/2}$ (note that $\bM$ is positive semi-definite so that its square root exists), and use the variable splitting trick:
\begin{align}\label{eq:DesignIntegratedSystem3}
\min_{\bs,t} & \ -t \nonumber \\
\textrm{s.t.}
& \ \by_k = {\bG}^\HH(\theta_k)\bs  - \bd_k, \|\by_k\|_2^2 \leq \varepsilon_k, k=1,\cdots,  N_0,\nonumber \\
&\ \bv = \bM_r \bs, \|\bv\|_2^2 \geq t,\nonumber \\
& \ \bs_n^\HH\bs_n = e_t/\NT, \text{PAPR}(\bs_n) \leq \rho, n = 1, \cdots, \NT.
\end{align}

The augmented Lagrangian associated with \eqref{eq:DesignIntegratedSystem3} is given by
\begin{align}
&L_{\mu}(\bs,t,\{\by_k\},\bv,\{\boldsymbol{\gamma}_k\},\blam) \nonumber \\
=& -t + \frac{\mu}{2} \left[\sum_{k=1}^{N_0}(\| \by_k - {\bG}^\HH(\theta_k)\bs  + \bd_k + \bgam_k\|_2^2 - \| \bgam_k\|_2^2)\right] \nonumber \\
&+\frac{\mu}{2} \left(\|\bv - \bM_r\bs + \blam\|_2^2 - \|\blam\|_2^2\right),
\end{align}
where $\mu$ is the penalty parameter.

The ADMM algorithm then consists of the following steps at the $(m+1)$-th iteration:
\begin{subequations}
  \begin{align}
\bs^{(m+1)}&= \arg\min_{\bs} L_\mu(\bs,t^{(m)},\{\by_k^{(m)}\},\bv^{(m)},\{\boldsymbol{\gamma}_k^{(m)}\},\blam^{(m)}) \label{eq:updates}\\
\by_k^{(m+1)}&= \arg\min_{\by_k} L_\mu(\bs^{(m+1)},t^{(m)},\by_k,\{\by_n^{(m+1)}\}_{n=1, n \neq k}^{N_0}, \bv^{(m)}, \{\boldsymbol{\gamma}_k^{(m)}\},\blam^{(m)}) \label{eq:updateyk}\\
\{\bv^{(m+1)}, t^{(m+1)}\}&= \arg\min_{\bv,t} L_\mu(\bs^{(m+1)},t,\{\by_k^{(m+1)}\},\bv,\{\boldsymbol{\gamma}_k^{(m)}\},\blam^{(m)}) \label{eq:updatev}\\
\bgam_{k}^{(m+1)}&= \bgam_{k}^{(m)} + \by_k^{(m+1)}-{\bG}^{\HH}(\theta_k) \bs^{(m+1)}  +\bd_k \\
\blam^{(m+1)}&=\bgam^{(m)} + \bv^{(m+1)}- \bM_r \bs^{(m+1)}
\end{align}
\end{subequations}

Next we present solutions to the optimization problems in  \eqref{eq:updates}, \eqref{eq:updateyk}, and \eqref{eq:updatev}. For simplicity, we omit the variable superscript if doing so leads to no confusion.

$\bullet$ Update of $\bs$:

Define
\begin{equation}\label{eq:matrixT}
  \bT = \sum_{k=1}^{N_0}{\bG}(\theta_k){\bG}^\HH(\theta_k) +\bM
\end{equation}
and
\begin{equation} \label{eq:vectort}
 \bt = \sum_{k=1}^{N_0}{\bG}(\theta_k) (\by_k + \bd_k + \bgam_{k}) + \bM_r(\bv+ \blam).
\end{equation}
Then \eqref{eq:updates} can be reformulated as:
\begin{align}\label{eq:SolveS}
  \min_{\bs} & \ \bs^\HH \bT \bs - 2 \Re(\bt^\HH \bs) \nonumber\\
  \textrm{s.t.} &\  \bs_n^\HH\bs_n = e_t/\NT, \text{PAPR}(\bs_n) \leq \rho, n = 1, \cdots, \NT.
\end{align}
The optimization problem in \eqref{eq:SolveS} can be tackled using the majorization-minimization (MM) method (see, e.g., \cite{Tang2021IT,Tang2021Profiling} for similar problems). To this end, we note that  $\bs^\HH \bT \bs$ is majorized by \cite[Equation (67)]{Tang2021IT} 
\begin{equation}
  2\Re(\bs^\HH \bar{\bT} \bs^{(t)}) + 2\lambda_{\max}(\bT)e_t - (\bs^{(t)})^\HH \bT \bs^{(t)},
\end{equation}
where $\bar{\bT} = \bT - \lambda_{\max}(\bT)\bI$, and we use the superscript $t$ to denote the $t$-th inner iteration. Therefore, the surrogate MM problem at the $(t+1)$th (inner) iteration is given by
\begin{align}\label{eq:SolveSmajorized}
  \min_{\bs} & \ - 2 \Re(\bar{\bt}^\HH \bs) \nonumber\\
  \textrm{s.t.} &\  \bs_n^\HH\bs_n = e_t/\NT, \text{PAPR}(\bs_n) \leq \rho, n = 1, \cdots, \NT,
\end{align}
where $\bar{\bt} = \bt - \bar{\bT} \bs^{(t)}$. The optimization problem in \eqref{eq:SolveSmajorized} can be split into $\NT$ independent problems, the $n$th of which is given by
\begin{align}\label{eq:SolvePAPR}
  \min_{\bs_n} & \ - 2 \Re(\bar{\bt}_n^\HH \bs_n) \nonumber\\
  \textrm{s.t.} &\  \bs_n^\HH\bs_n = e_t/\NT, \text{PAPR}(\bs_n) \leq \rho,
\end{align}
where $\bar{t}_n(l)$, i.e., the $l$th element of $\bar{\bt}_n$, satisfies $\bar{t}_n(l) = \bar{\bt}(n+(l-1)\NT)$. The problem in \eqref{eq:SolvePAPR} can be solved by Algorithm 2 in \cite{Tropp2005AP}. Moreover, if $\rho=1$, the solution is simply given by
\begin{equation}
  s_n(l) = a_s \exp(j\arg(\bar{t}_n(l))).
\end{equation}
Algorithm \ref{Alg:MM} summarizes the procedure for solving problem \eqref{eq:SolveS}; we terminate the algorithm if the relative change of the objective is smaller than a predefined threshold. Note that one  can employ the FISTA method \cite{Beck2009FISTA} or the SQUAREM method \cite{varadhan2008SQUAREM} to accelerate Algorithm \ref{Alg:MM}.
\begin{algorithm}[!htp]
  \caption{ \small  MM algorithm for the problem in \eqref{eq:SolveS}.}\label{Alg:MM}
  \KwIn{$\bT, \bt$.}
  \KwOut{$\bs^{(m+1)}$.}
  \textbf{Initialize:} $\bs^{(m,0)}$ (e.g., $\bs^{(m,0)} = \bs^{(m)}$), $\bar{\bT} = \bT - \lambda_{\max}(\bT)\bI$. \\
    \Repeat{convergence}{
    $\bar{\bt} = \bt - \bar{\bT} \bs^{(m,t)}$.\\
    Obtain $\bs^{(m,t+1)}$ by solving the problem in \eqref{eq:SolvePAPR}.\\
    $t= t+ 1$.\\
    }
    $\bs^{(m+1)}$ = $\bs^{(m,t)}$.
\end{algorithm}

$\bullet$ Update of $\by_k$:

Let $\bz_k = {\bG}^\HH(\theta_k)\bs  - \bd - \bgam_k$. Then the optimization problem in \eqref{eq:updateyk} can be written as
\begin{align}\label{eq:Designy1}
  \min_{\by_k} & \ \|\by_k - \bz_k\|_2^2 \nonumber\\
   \textrm{s.t.} & \ \|\by_k\|_2^2 \leq \varepsilon_k.
\end{align}

It can be readily checked that the solution to \eqref{eq:Designy1} is given by
\begin{equation}
  \by_k = \begin{cases}
            \bz_k, & \mbox{if } \|\bz_k\|_2^2 \leq \varepsilon_k,\\
            \sqrt{\varepsilon_k}\frac{\bz_k}{\|\bz_k\|_2}, & \mbox{otherwise}.
          \end{cases}
\end{equation}

$\bullet$ Update of $\{\bv,t\}$:

Let $\bar{\bz} = \bM_r\bs - \blam$. Then the optimization problem in \eqref{eq:updatev} can be written as
\begin{align}\label{eq:Designy2}
  \min_{\bv,t} & \ \frac{\mu}{2}\|\bv - \bar{\bz}\|_2^2 - t \nonumber\\
   \textrm{s.t.} & \ \|\bv\|_2^2 \geq t.
\end{align}

For fixed $\bv$, the minimizer $t$ of  \eqref{eq:Designy2} is
\begin{equation} \label{eq:tOPT}
  t = \|\bv\|_2^2.
\end{equation}
Inserting  \eqref{eq:tOPT} into \eqref{eq:Designy2}  leads to the following problem in $\bv$:
\begin{align}
  \min_{\bv} & \ \frac{\mu}{2}\|\bv - \bar{\bz}\|_2^2 - \|\bv\|_2^2.
\end{align}
For $\mu>2$ (which we assume), the minimizer of the above problem is given by
\begin{equation}
  \bv = \frac{\mu \bar{\bz}}{\mu-2}.
\end{equation}

We summarize the steps of the proposed ADMM algorithm in Algorithm \ref{Alg:ADMM}, in which we stop the iterations if the change of the SINR is insignificant.

\textit{Remark 4:} In the formulation of \eqref{eq:SigDesign}, the values of the matching errors (i.e., $\varepsilon_k ,k=1, \cdots, N_0$) should be set according to practical considerations (e.g., the allowed matching error for communication signals is usually smaller than that for jamming signals). Note that these values are set too small, the optimization problem in \eqref{eq:SigDesign} may be infeasible. In such a situation, we should increase  the values of $\varepsilon_k$ $(k=1, \cdots, N_0)$ to make  the problem feasible.

\begin{algorithm}[!htp]
  \caption{ \small  ADMM algorithm for the problem in \eqref{eq:SigDesign}.}\label{Alg:ADMM}
  \KwIn{$\bR, \theta_\textrm{t}, \{\theta_k\}_{k=1}^{N_0}, \{\bd_k\}_{k=1}^{N_0} ,e_t,\rho$.}
  \KwOut{$\bs^{*}$.}
  \textbf{Initialize:} \\
  $\mu$, $\{\by_k^{(0)}\}_{k=1}^{N_0} = \bzero, \bv^{(0)} =\bzero, \{\bgam_k^{(0)}\}_{k=1}^{N_0} = \bzero, \blam = \bzero$; \\ Compute $\bT$ and its largest eigenvalue. \\
  $m=0$.\\
    \Repeat{convergence}{
    \tcp*[h]{\textit{\textrm{Update of}} $\bs^{(m+1)}$ }\\
    Compute $\bt$ using \eqref{eq:vectort}.\\
    Update $\bs^{(m+1)}$ via Algorithm \ref{Alg:MM}.\\
    \tcp*[h]{\textrm{\textit{Update of }}$\by_k^{(m+1)}$ }\\
    \For{$k=1,2,\cdots,N_0$}{
    $\bz_k = {\bG}^\HH(\theta_k)\bs^{(m+1)}  - \bd_k - \bgam_k^{(m)}$. \\
    $\by_k^{(m+1)} = \min({\sqrt{\varepsilon_k}}/{\|\bz_k\|_2},1)\cdot \bz_k$.\\}
    \tcp*[h]{\textit{\textrm{Update of }}$\{\bv^{(m+1)},t^{(m+1)}\}$ }\\
    $\bar{\bz} = \bM_r\bs^{(m+1)} - \blam^{(m)}$.\\
    $\bv^{(m+1)}= \frac{\mu \bar{\bz}}{\mu-2}$.\\
    $t^{(m+1)}= \|\bv^{(m+1)}\|_2^2$. \\
    \tcp*[h]{\textit{\textrm{Update of }}$\bgam_{k}^{(m+1)}$ }\\
    $\bgam_{k}^{(m+1)} = \bgam_{k}^{(m)} + \by_k^{(m+1)} -{\bG}^{\HH}(\theta_k) \bs^{(m+1)}  +\bd_k.$ \\
    \tcp*[h]{\textit{\textrm{Update of }}$\blam^{(m+1)}$ }\\
    $\blam^{(m+1)} =\blam^{(m)} + \bv^{(m+1)} - \bM_r \bs^{(m+1)} .$ \\
    $m=m+1$.
    }
    $\bs^{\star}$ = $\bs^{(m)}$.
\end{algorithm}

\section{Analysis and Discussions} \label{sec:Discussion}
\subsection{Computational Complexity Analysis}
In this section, we analyze the per-iteration computational complexity of the proposed ADMM algorithm. Table \ref{Table:2} summarizes the computational complexity associated with each step of  Algorithm \ref{Alg:ADMM}, where $N_{\textrm{papr}}$ is the number of iterations needed to reach convergence for Algorithm \ref{Alg:MM}, and we note that  the update of $\bgam_{k}^{(m+1)}$ and $\blam^{(m+1)}$ can share the matrix multiplication result in the computation of $\bz_k$ and $\bar{\bz}$, respectively. In addition, we have used the fact  that for any matrix $\bX = \bI \otimes \bar{\bX}$, the multiplication of $\bX$ with a vector $\bq$ of appropriate size can be efficiently done as follows:
\begin{equation}
  \bX \bq = (\bI \otimes \bar{\bX}) \bq = \vec(\bar{\bX} \bQ),
\end{equation}
where $\bq = \vec(\bQ)$.
Therefore, the per-iteration complexity of the proposed algorithm is $O(N_0\NT L + N_{\textrm{papr}}N_\textrm{T}^2L^2)$. Note that if $\bR$ has the structure discussed in Section \ref{sec:structured}, then the computational complexity of the proposed algorithm can be further reduced. Specifically, in such a case, $\bT$ can be written as%
\begin{align}
  \bT
  &= \bI_L \otimes \left[\sum_{n=1}^{N_0}\ba(\theta_k)\ba^\HH(\theta_k) + \textrm{SINR}_{{\textrm{R}}} \ba(\theta_\textrm{t})\ba^\HH(\theta_\textrm{t})\right]. 
\end{align}
Using this observation, the complexity of updating $\bs^{(m+1)}$ is $O(N_{\textrm{papr}}N_\textrm{T}^2L)$, and the per-iteration computational complexity of the proposed algorithm can be reduced to $O(N_0\NT L + N_{\textrm{papr}}N_\textrm{T}^2 L)$.

\begin{table}[!htbp]
\centering
\caption{Computational complexity Per-iteration}
\begin{tabular}{|c|c|}
\hline
  Computation & Complexity \\
  \hline
   $\bt$ & $O(N_0\NT L)$ \\
   \hline
   $\bs^{(m+1)}$ &$O(N_{\textrm{papr}}N_\textrm{T}^2L^2)$\\
   \hline
   $\{\bz_k\}_{k=1}^{N_0}$ & $O(N_0\NT L)$ \\
   \hline
   $\{\by_k^{(m+1)}\}_{k=1}^{N_0}$ & $O(N_0 L)$\\
   \hline
   $\bar{\bz} $ & $O(N_\textrm{T}^2 L^2)$ \\
   \hline
   $\bv^{(m+1)}$ & $O(N_\textrm{T} L)$\\
   \hline
   $t^{(m+1)}$ & $O(N_\textrm{T} L)$ \\
  \hline
\end{tabular}\label{Table:2}
\end{table}
\subsection{The Achievable Rate}
Assume that $L \gg 1$. Then according to the law of large numbers, we can write:
\begin{align*}
  &\|\ba^\HH(\theta_{\textrm{c},n})\bS-\bd_{\textrm{c},n}\|_2^2 \\
  &\approx L\mathbb{E}[|\ba^\HH(\theta_{\textrm{c},n})\bS(:,k)-d_{\textrm{c},n}(k)|^2] \leq \varepsilon_n, n=1,\cdots,N_{\textrm{c}}.
\end{align*}
As a result, 
\begin{equation}
  \chi_n \geq \frac{\textrm{CSNR}}{1 + \varepsilon_n/L},
\end{equation}
and the achievable sum rate satisfies (see \textit{Remark} 2 in Section \ref{sec:SigModel})
\begin{equation}
  R \geq \sum_{n=1}^{N_\textrm{c}} {\log_2 \left(1+\frac{\textrm{CSNR}}{1 + \varepsilon_n/L}\right)}.
\end{equation}
\subsection{The Jamming Power}
Let $\by_{\textrm{jam},n} = \bS^\top \ba^*(\theta_{n})$ be the emitted jamming waveform in the direction $\theta_{n}$, $n = N_\textrm{c}+1,N_\textrm{c}+2,\cdots,N_0$. Note that
\begin{equation}
  \left(\|\by_{\textrm{jam},n} \|_2 - \|\bd_n\|_2\right)^2 \leq \| \by_{\textrm{jam},n} - \bd_n\|_2^2 \leq \varepsilon_n.
\end{equation}
Therefore, the jamming power at $\theta_{n}$ satisfies (assuming that $\|\bd_n\|_2^2 \gg \varepsilon_n$)
\begin{equation}
  (\|\bd_n\|_2 - \sqrt{\varepsilon_n})^2 \leq \|\by_{\textrm{jam},n} \|_2^2 \leq (\|\bd_n\|_2 + \sqrt{\varepsilon_n})^2.
\end{equation}

\section{Numerical Examples} \label{sec:NumericalExamples}
In this section, we demonstrate the results and the performance of the proposed algorithm through several numerical examples. We consider a MIMO MFRF system with $\NT=12$ transmit antennas and $\NR=12$ receive antennas. Both antenna arrays are uniform linear arrays. The inter-element spacings of the transmit array and the receive array are $d_\txT = d_\txR = \lambda/2$, where $\lambda$ is the wavelength. The target direction is $\theta_\textrm{t} = 0^\circ$. The disturbance covariance matrix $\bR$ has the Kronecker-product structure discussed in Section \ref{sec:structured}. For such a structure, the receive SINR  does not depend on the waveform matrix $\bS$. Thus, unless otherwise stated, the SINR mentioned in this section is the transmit SINR.  Finally, all the numerical simulations are performed on a standard laptop with CPU CoRe i7-8550U and 8 GB memory.
\subsection{Performance of Energy-Constrained Waveforms}
\begin{figure}[!htp]
\centering
{\subfigure[]{{\includegraphics[width = 0.45\textwidth]{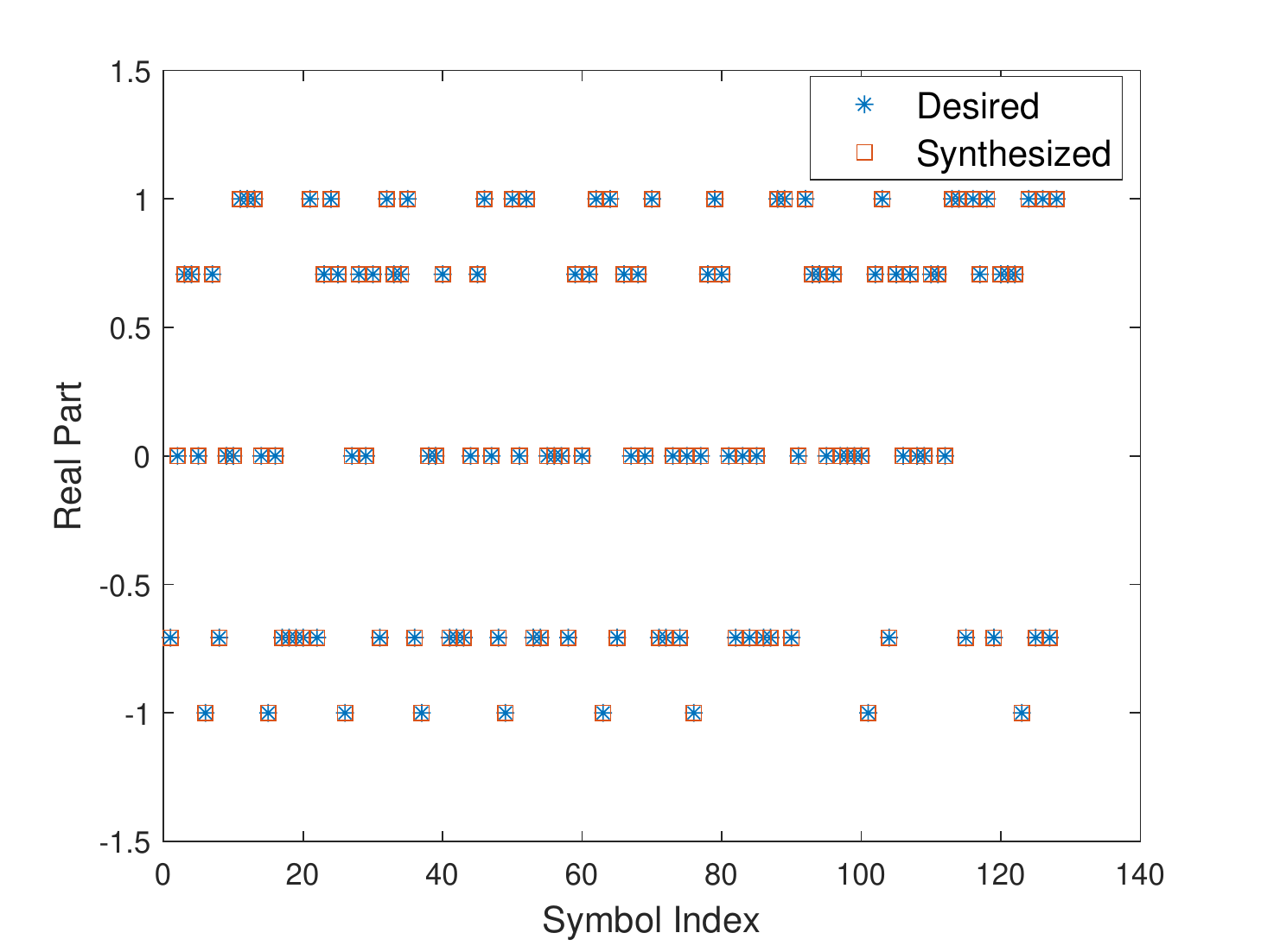}} \label{Fig:2a}} }
{\subfigure[]{{\includegraphics[width = 0.45\textwidth]{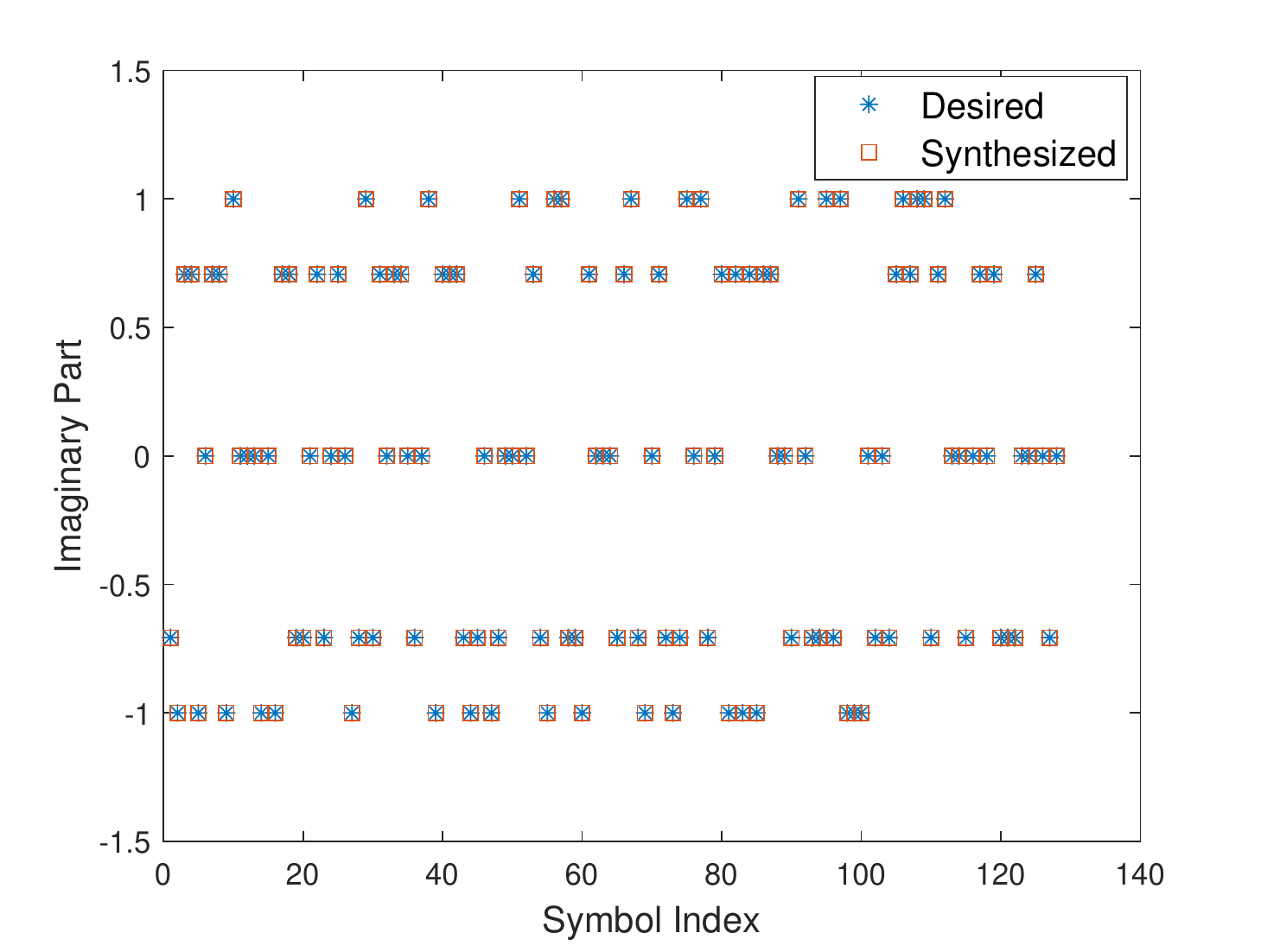}} \label{Fig:2b}} }
{\subfigure[]{{\includegraphics[width = 0.45\textwidth]{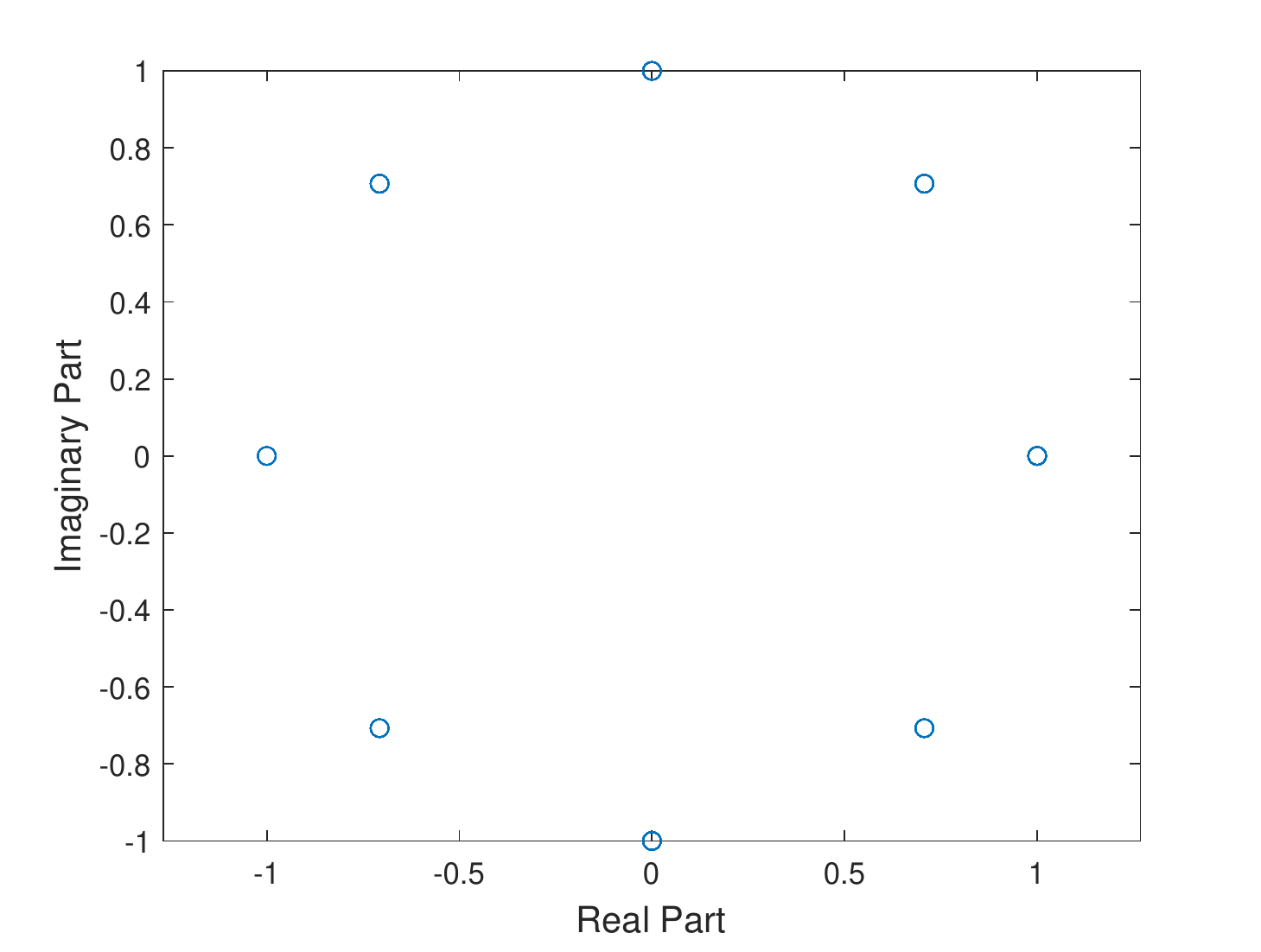}} \label{Fig:2c}} }
\caption{The synthesized 8PSK signals at $\theta_\textrm{c} = -25^\circ$.  (a) The real part. (b) The imaginary part. (c) The constellation diagram.}
\label{Fig:2}
\end{figure}

\begin{figure*}[!htp]
\centering
{\subfigure[]{{\includegraphics[width = 0.45\textwidth]{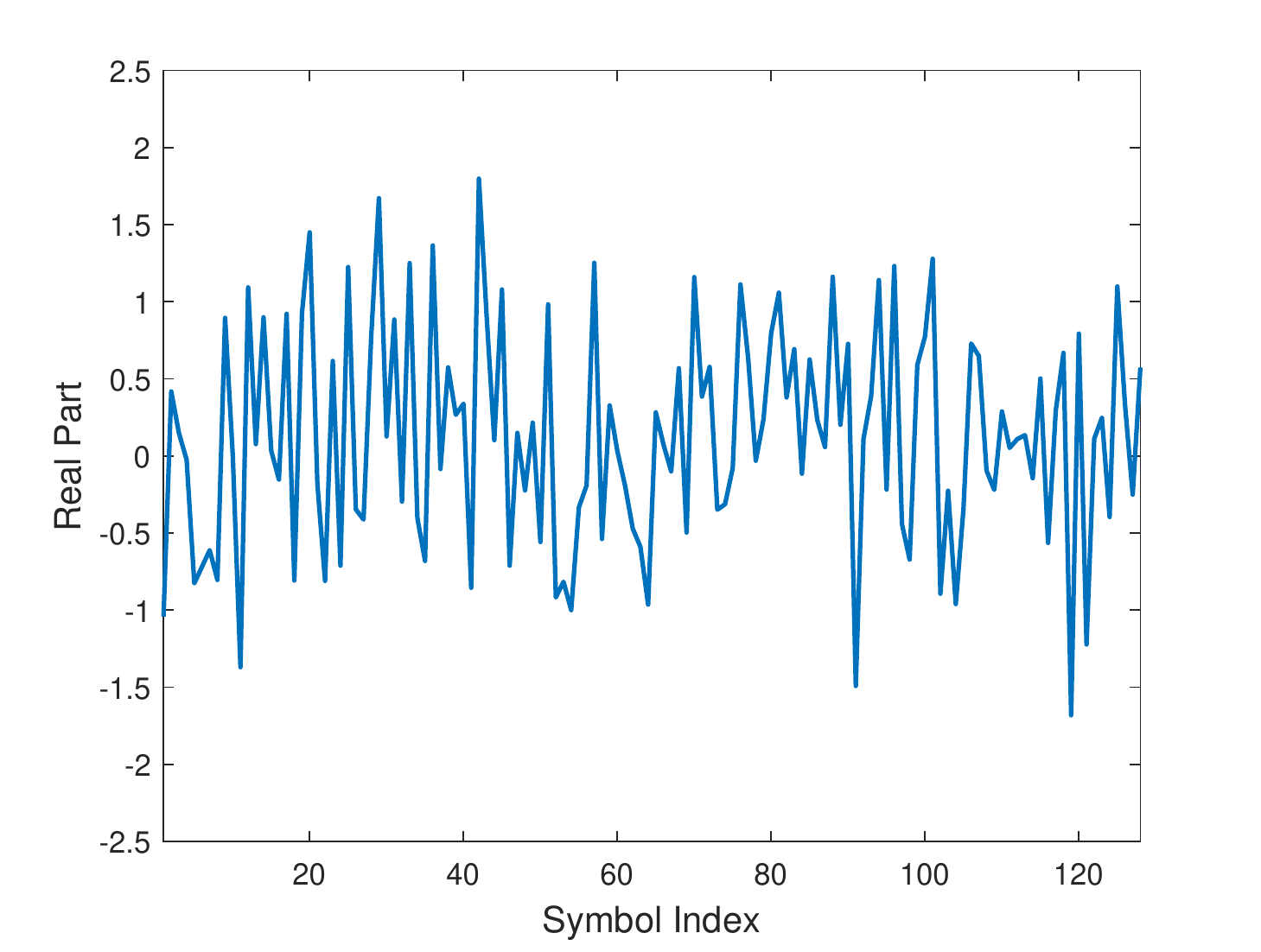}} \label{Fig:3a}} }
{\subfigure[]{{\includegraphics[width = 0.45\textwidth]{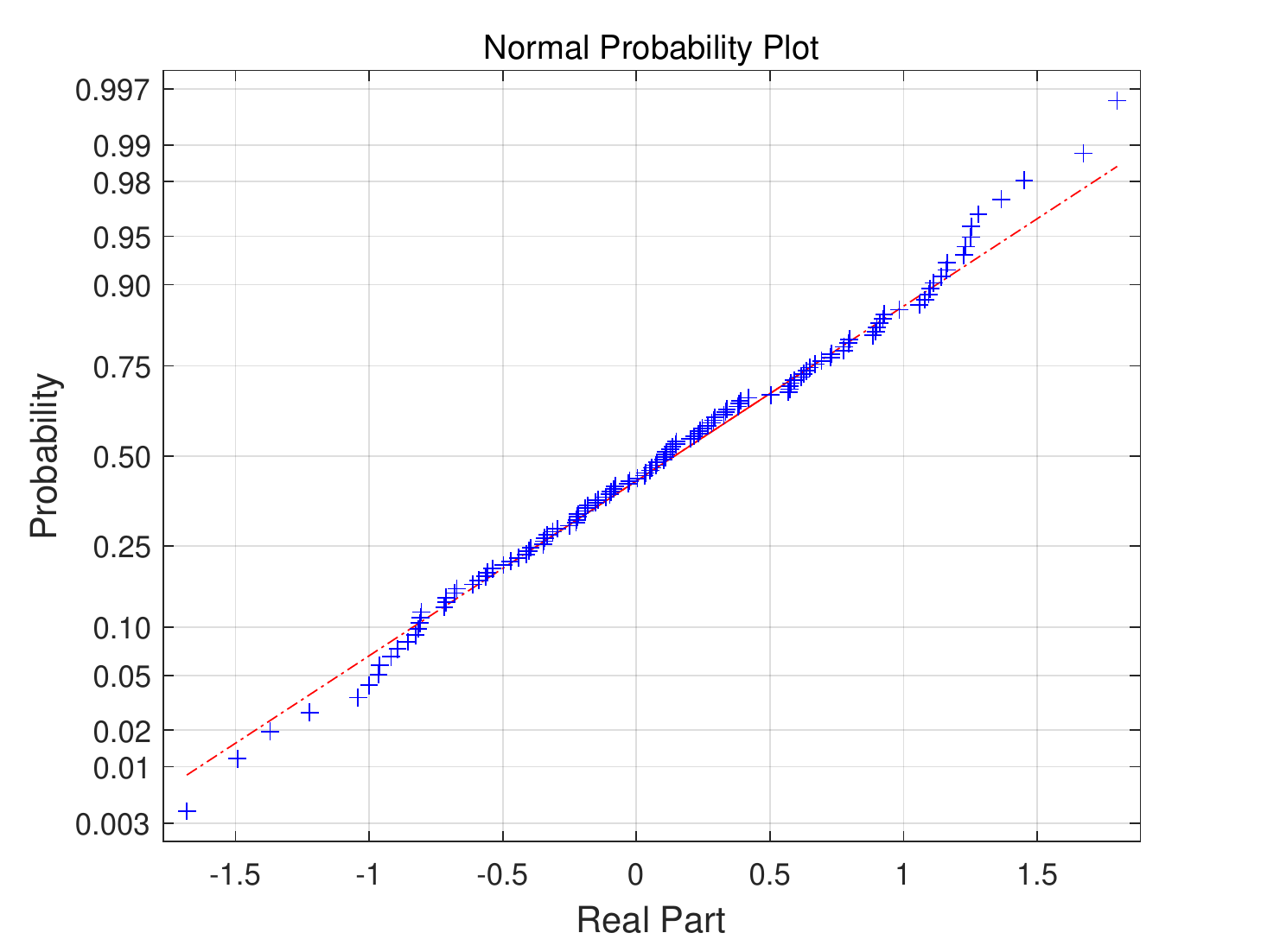}} \label{Fig:3b}} }
{\subfigure[]{{\includegraphics[width = 0.45\textwidth]{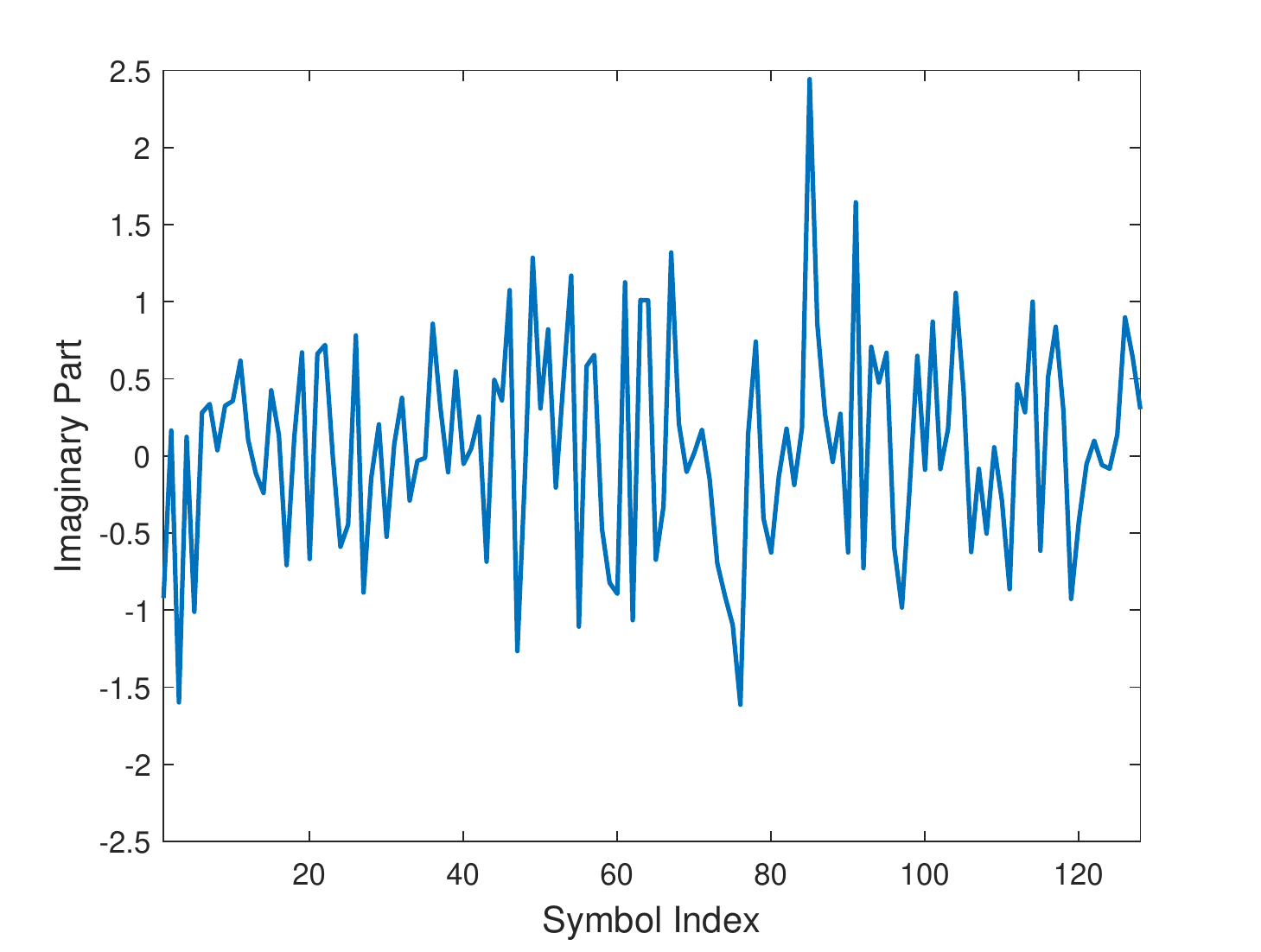}} \label{Fig:3c}} }
{\subfigure[]{{\includegraphics[width = 0.45\textwidth]{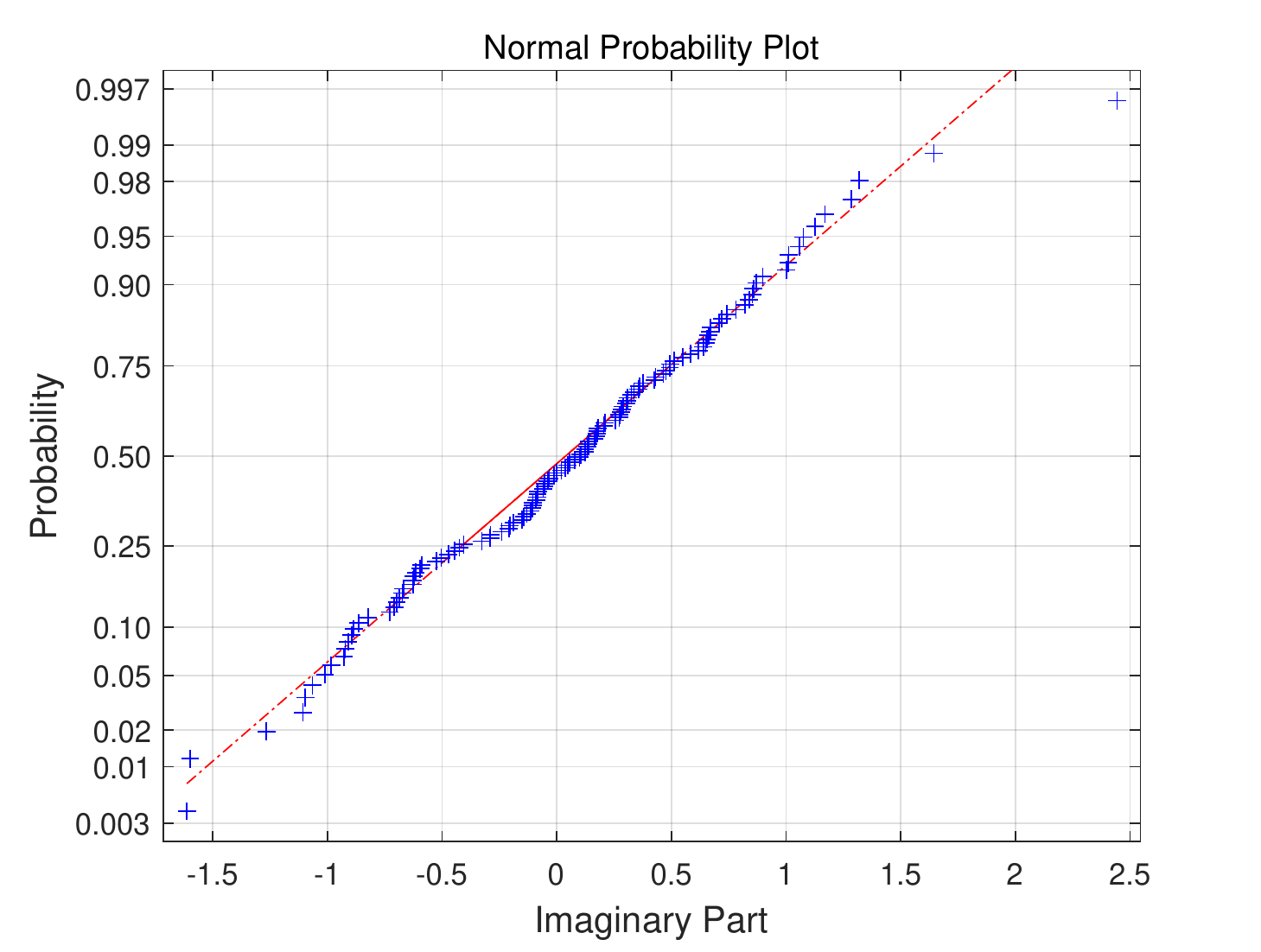}} \label{Fig:3d}} }
\caption{The synthesized noise-like signals at $\theta_{\textrm{jam}} = 20^\circ$.  (a) The real part. (b) The normal probability plot of the real part. (c) The imaginary part. (d) The normal probability plot of the imaginary part.}
\label{Fig:3}
\end{figure*}

First, we assume that the directions of the communication receiver and the hostile target are $\theta_\textrm{c} = -25^\circ$ and  $\theta_{\textrm{jam}} = 20^\circ$, respectively. We want to send a 8-phase-shift keying (PSK) signal to the communication receiver and a noise-like jamming signal to the hostile target. The amplitude of the desired PSK signal is 1 and the symbol bits are randomly generated. The noise-like signal is randomly generated with elements drawn from a  Gaussian distribution with zero mean and variance of 1. Fig. \ref{Fig:2a} and Fig. \ref{Fig:2b} show the real part and the imaginary part of the synthesized signals at $\theta_\textrm{c} = -25^\circ$, for a code length $L=128$, and transmit energy $e_t = 4L/\NT$. We can see that the synthesized signals and the desired ones overlap with each other. Moreover, Fig. \ref{Fig:2c} shows that the synthesized signal has a distortionless  constellation diagram. Fig. \ref{Fig:3a} and Fig. \ref{Fig:3c} plot the synthesized signals at $\theta_{\textrm{jam}} = 20^\circ$. The noise-like property of these signals is confirmed by the normal probability plots in Fig. \ref{Fig:3b} and Fig. \ref{Fig:3d}. The results in these figures indicate that the synthesized signals follow a normal distribution as desired. In sum, the MIMO MFRF system is able to simultaneously transmit communication signals to a legitimate receiver and noise-like signals to a hostile target.

\begin{figure*}[!htp]
\centering
{\subfigure[]{{\includegraphics[width = 0.45\textwidth]{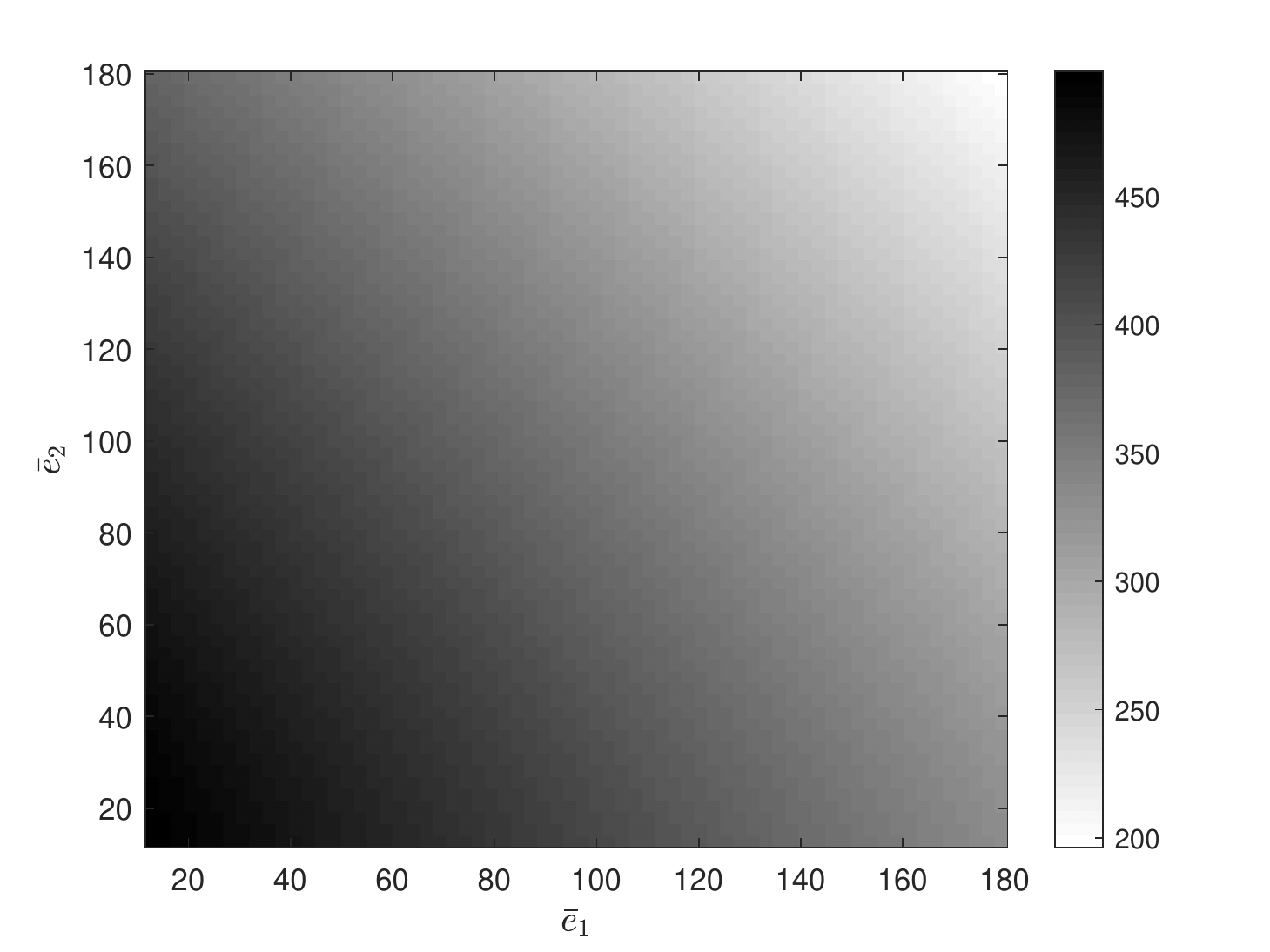}} \label{Fig:4a}} }
{\subfigure[]{{\includegraphics[width = 0.45\textwidth]{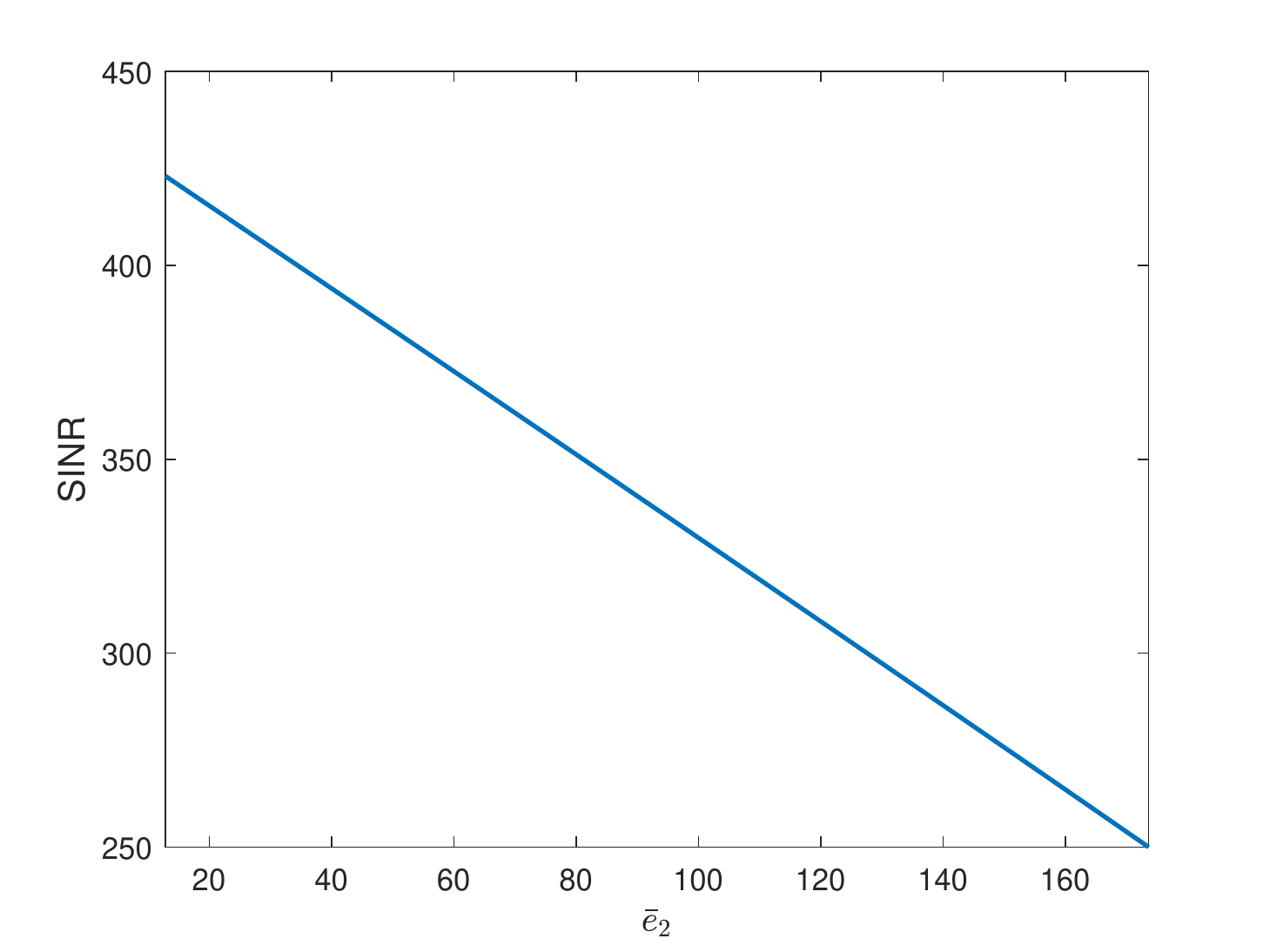}} \label{Fig:4b}} }
{\subfigure[]{{\includegraphics[width = 0.45\textwidth]{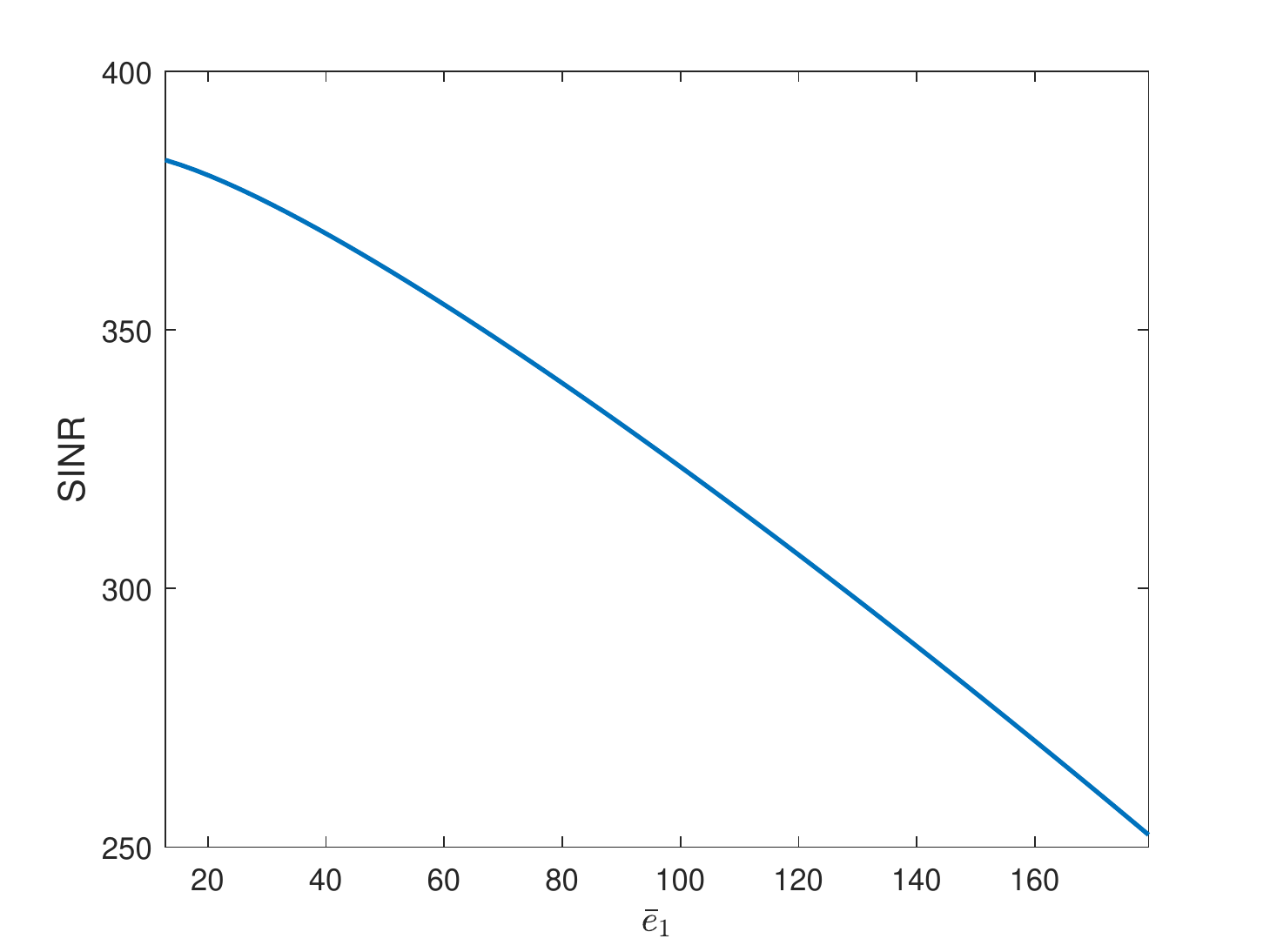}} \label{Fig:4c}} }
{\subfigure[]{{\includegraphics[width = 0.45\textwidth]{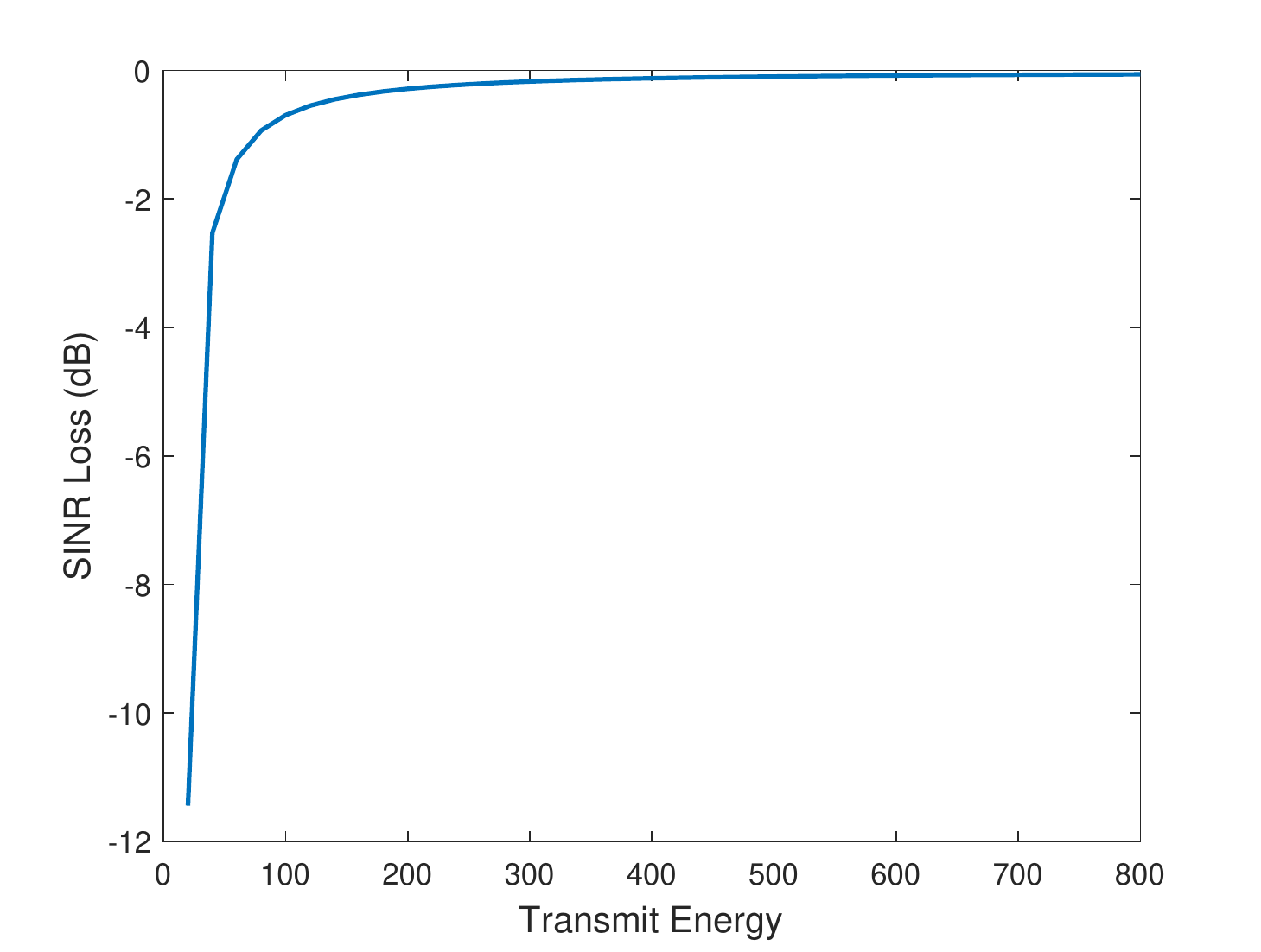}} \label{Fig:4d}} }
\caption{The SINR of the MFRF system. (a) The SINR versus the transmit energy of communication signals (i.e., $\bar{e}_1$) and noise-like signals (i.e., $\bar{e}_2$). $e_t = 4L/\NT$. (b) The cut of Fig. \ref{Fig:4a} at $\bar{e}_1 = 128$. (c)  The cut of Fig. \ref{Fig:4a} at $\bar{e}_2= 128$. (d) The SINR loss compared with a radar-only system.}
\label{Fig:4}
\end{figure*}

Fig. \ref{Fig:4a} shows the SINR of the MFRF system versus the transmit energy of the desired communication signals and noise-like jamming signals. We use the same parameter setting as for Fig. \ref{Fig:2} and Fig. \ref{Fig:3}, but vary the amplitude of the transmitted waveforms to satisfy the energy constraint. Fig. \ref{Fig:4b} and Fig. \ref{Fig:4c} show the cut of Fig. \ref{Fig:4a} at $\bar{e}_1 = 128$ and $\bar{e}_2= 128$, respectively. One can see that the increase of the transmit energy of the desired communication signals and noise-like signals decreases the SINR. Recall the dependence of the achievable sum-rate and the ECM capability on the transmit energy of the two kinds of signals (see Section \ref{sec:structured}). As expected, there is a tradeoff between the SINR and the transmit energy of the desired communication signals and noise-like signals. Fig. \ref{Fig:4d} presents the SINR loss of the MFRF system compared with the radar-only system, versus the transmit energy of the system (i.e., $e_t$), where the transmit energy of the desired communication signals and noise-like signals is fixed (the same as for Fig. \ref{Fig:2} and Fig. \ref{Fig:3}). The figure shows that increasing the transmit energy of the system reduces the SINR loss, which is consistent with the result in Section \ref{Subsec:achievableSINR}.

\begin{figure}[!htp]
\centering
\includegraphics[width = 0.45\textwidth]{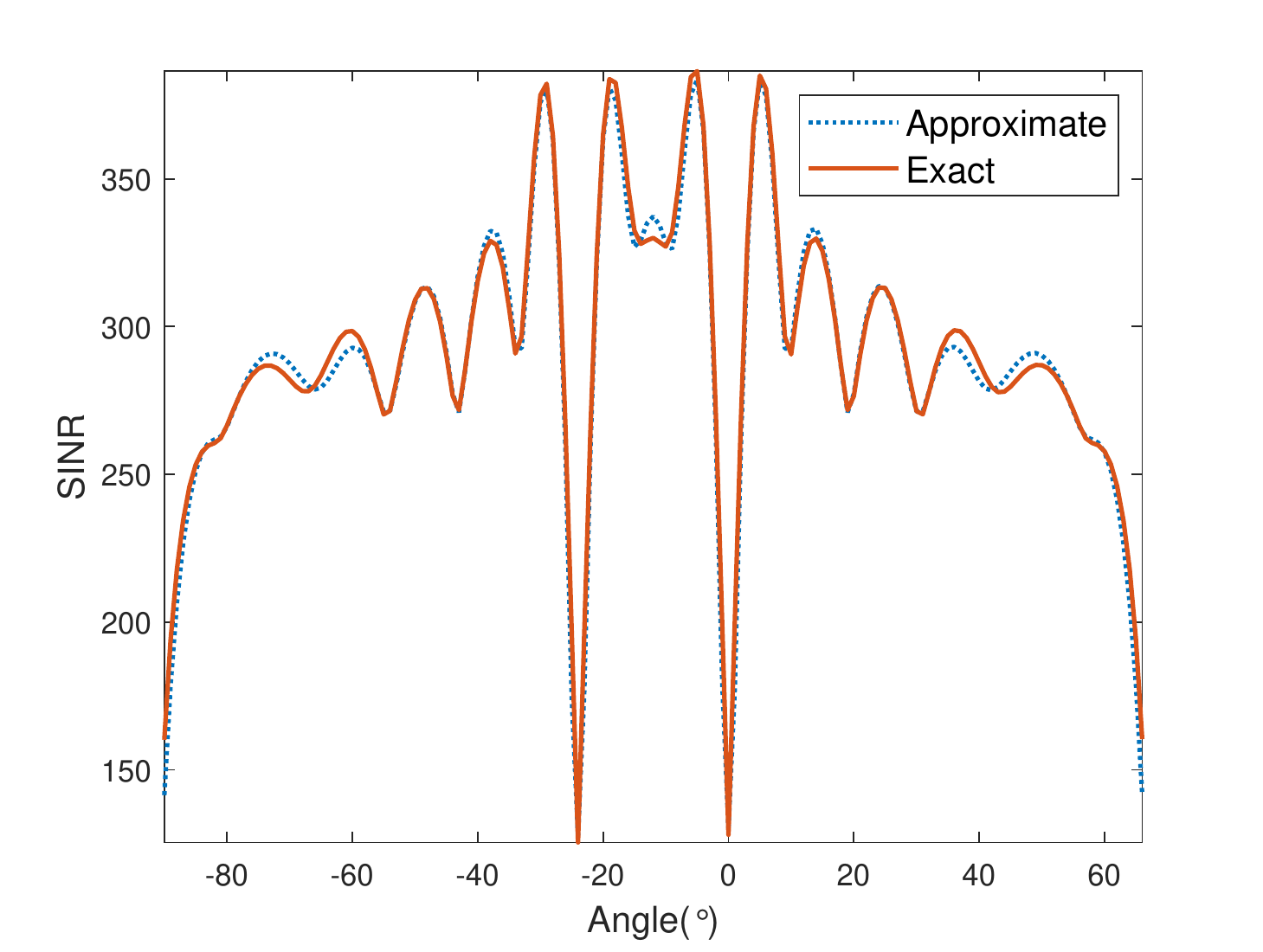}
\caption{The achievable SINR of the MFRF system. $L=128$. $e_t = 4L/\NT$.}
\label{Fig:5}
\end{figure}

Next we assess the impact of the locations of the communication receiver and the hostile target on the achievable SINR of the MFRF system. Fig. \ref{Fig:5} shows the transmit SINR of the system (including both the exact expression in \eqref{eq:SINRT} and the approximation in  \eqref{eq:SINRApprox}) versus the direction of the communication receiver, where we assume that the directions of the communication receiver and the hostile target satisfy $\theta_{\textrm{jam}}  - \theta_\textrm{c} =24 ^\circ$; the other parameter settings are the same as in Fig. \ref{Fig:2} and Fig. \ref{Fig:3}. We observe that there is a good agreement between the exact SINR and the approximate one, implying that the approximations in Section \ref{Subsec:achievableSINR} can provide a simple and useful guideline for analyzing the performance of MFRF systems. Also note that, the achievable SINR varies significantly with the directions of the communication receiver and the hostile target. For the present parameter setting, if the direction of the communication receiver or the hostile target is the same as that of the target, then the achievable SINR is the lowest (i.e., $\textrm{SINR}_\txT \approx \bar{e}_1 \approx \bar{e}_2 \approx L$, indicating an SINR loss of $6$ dB). On the other hand, for $\theta_\textrm{c} $ of $-29^\circ$, $-19^\circ$, $-5^\circ$, or $5^\circ$, the achievable SINR reaches its highest value (note that $\bar{e}_1 \approx \bar{e}_2$ and for these angles $G_{\textrm{sos}}\approx {\bar{e}}/(\NT e_t -\bar{e})$). Specifically, the highest SINR is about $e_t\NT - (N_0-1)\bar{e} \approx 3L$, with a corresponding SINR loss of $1.25$ dB.

Finally, we investigate the performance of the MFRF system when the numbers of communication receivers and hostile targets are varied. Table \ref{Table:3} shows the SINR of the system with the transmit energy $e_t = 500$ and $L=128$, for different numbers of communication receivers and hostile targets. The desired communication signals and jamming signals are generated like in Fig. \ref{Fig:2} and Fig \ref{Fig:3}. The directions of these  communication receivers and hostile targets  are listed in Table \ref{Table:3}. The SINR decreases with the number of communication receivers and hostile targets, which was as expected and is also consistent with the analysis in Section \ref{Subsec:achievableSINR}.%

\begin{table}[!htbp]
\centering
\caption{The SINR of the MFRF system for different numbers of communication receivers and hostile targets}
\begin{tabular}{|c|c|c|c|}
\hline
  &$\bTheta_{\textrm{c}}$& $\bTheta_{\textrm{jam}}$& SINR (dB) \\
  \hline
$N_\textrm{c} = 1,N_\textrm{jam} = 1$ & -25$^\circ$ & 20$^\circ$& 37.69\\
\hline
$N_\textrm{c} = 2,N_\textrm{jam} = 1$ & -25$^\circ$, -30$^\circ$ & 20$^\circ$& 37.57\\
\hline
$N_\textrm{c} = 2,N_\textrm{jam} = 2$ & -25$^\circ$, -30$^\circ$ & 20$^\circ$, 25$^\circ$& 37.33\\
\hline
$N_\textrm{c} = 3,N_\textrm{jam} = 2$ & -25$^\circ$, -30$^\circ$, -35$^\circ$ & 20$^\circ$, 25$^\circ$& 35.42\\
\hline
$N_\textrm{c} = 3,N_\textrm{jam} = 3$ & -25$^\circ$, -30$^\circ$, -35$^\circ$ & 20$^\circ$, 25$^\circ$, 30$^\circ$& 33.47\\
  \hline
\end{tabular}\label{Table:3}
\end{table}

\subsection{Performance of Constant-Modulus Waveforms}
\begin{figure}[!htp]
\centering
\includegraphics[width = 0.45\textwidth]{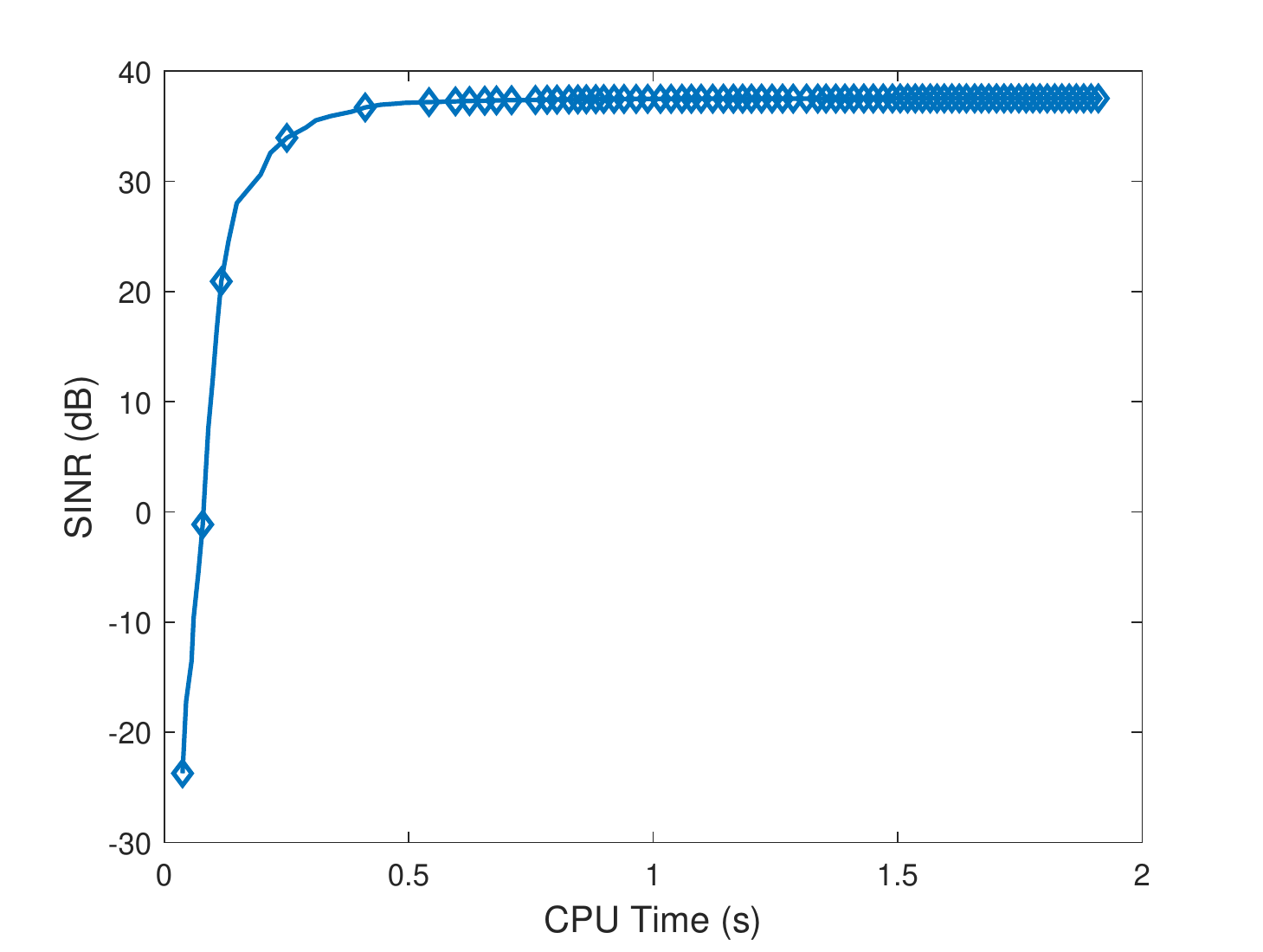}
\caption{The convergence of the proposed algorithm. $L=128$. $e_t = 500$. $\rho = 1$.}
\label{Fig:6}
\end{figure}

\begin{figure}[!htp]
\centering
{\subfigure[]{{\includegraphics[width = 0.45\textwidth]{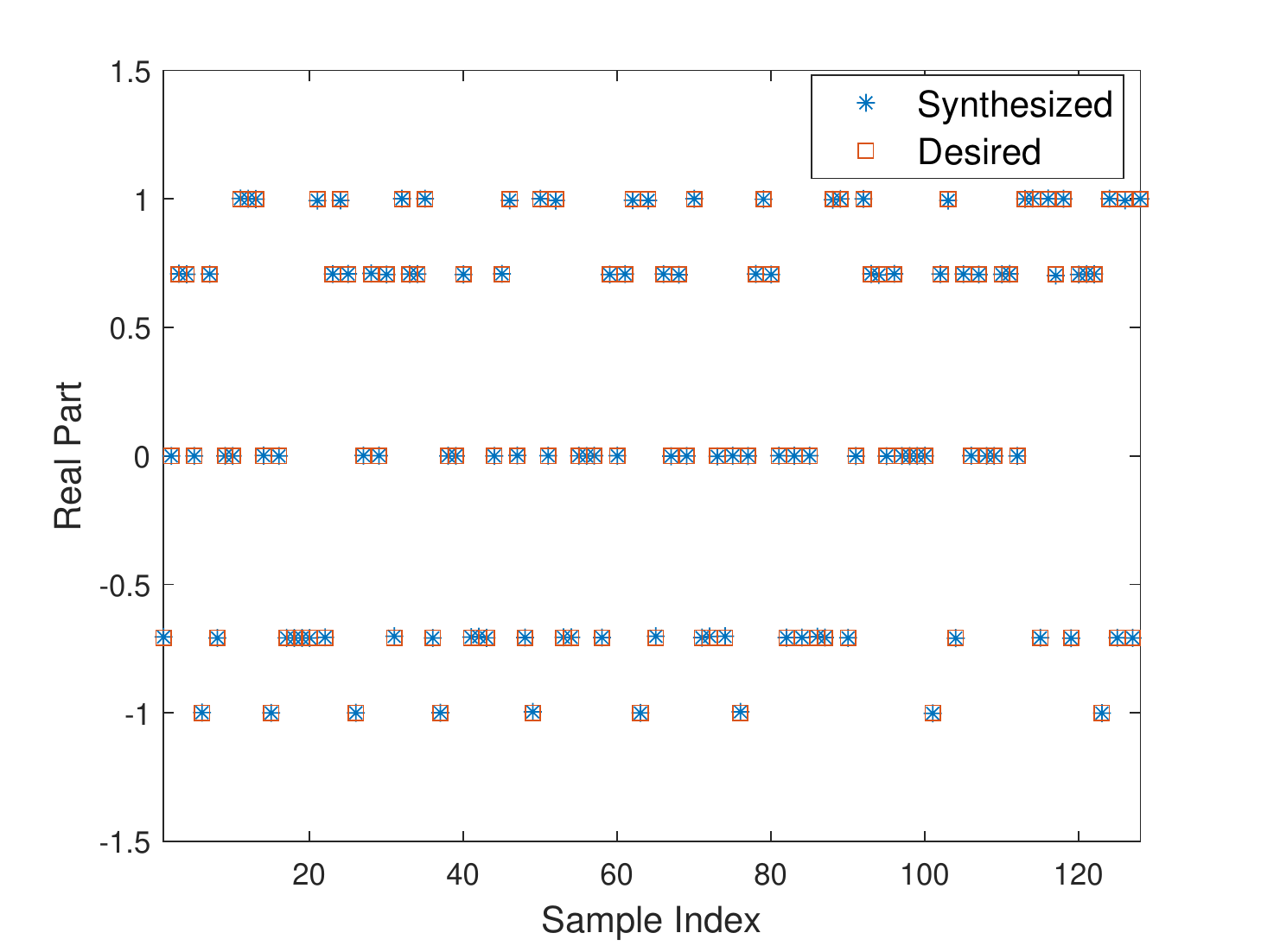}} \label{Fig:7a}} }
{\subfigure[]{{\includegraphics[width = 0.45\textwidth]{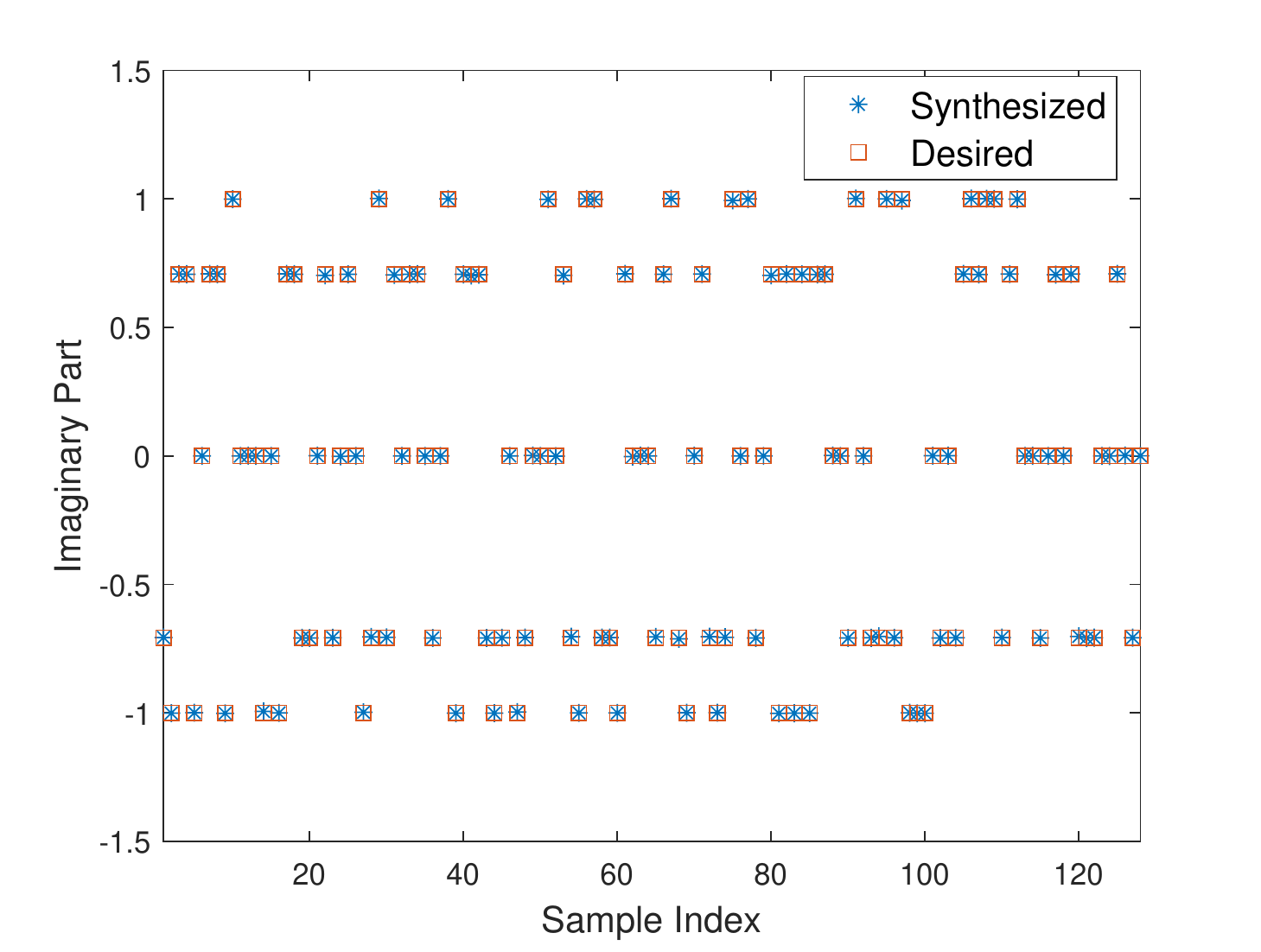}} \label{Fig:7b}} }
{\subfigure[]{{\includegraphics[width = 0.45\textwidth]{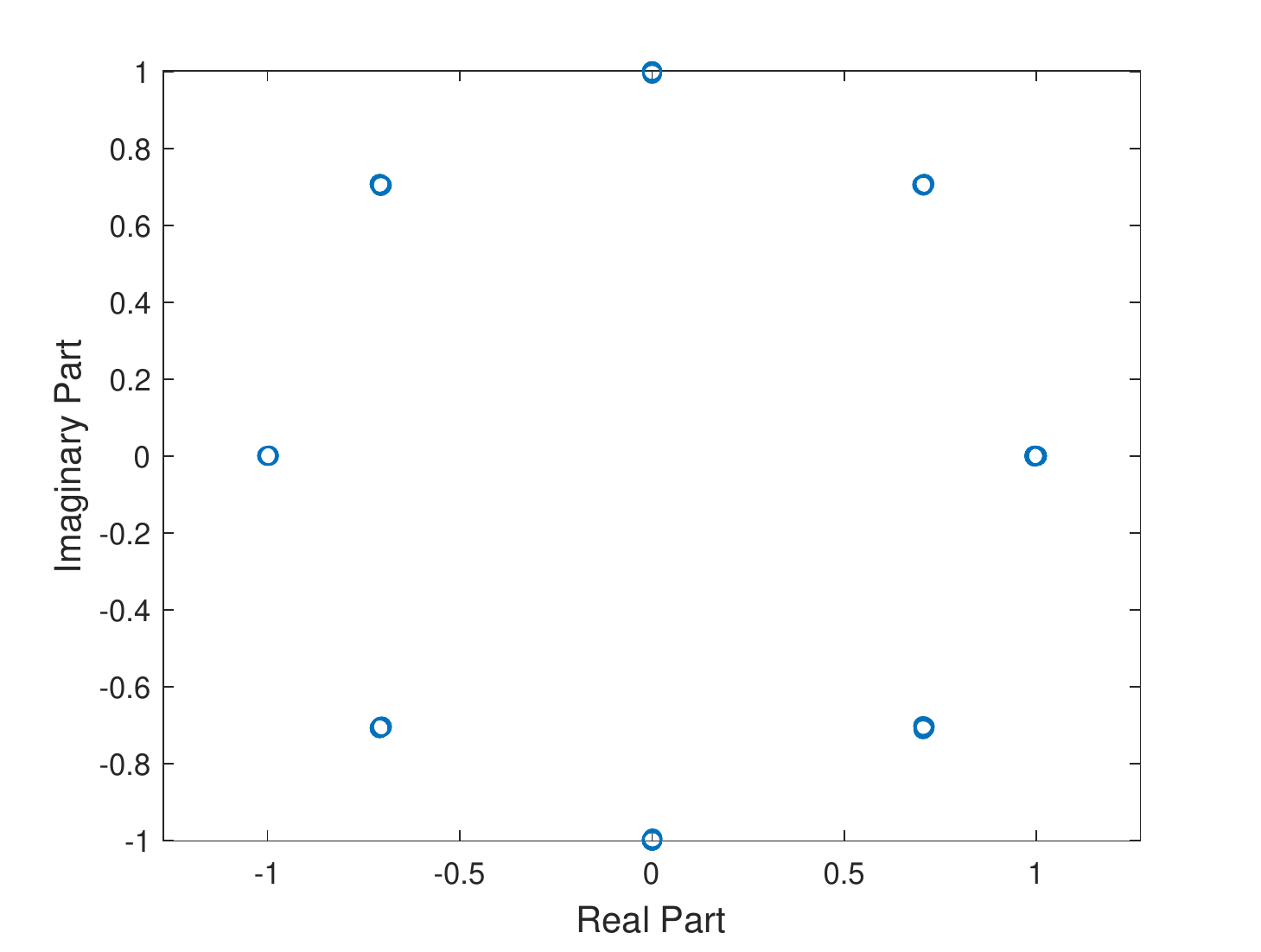}} \label{Fig:7c}} }
\caption{The synthesized 8PSK signals at $\theta_\textrm{c} = -25^\circ$.  $\rho = 1$. (a) The real part. (b) The imaginary part. (c) The constellation diagram.}
\label{Fig:7}
\end{figure}

\begin{figure}[!htp]
\centering
{\subfigure[]{{\includegraphics[width = 0.45\textwidth]{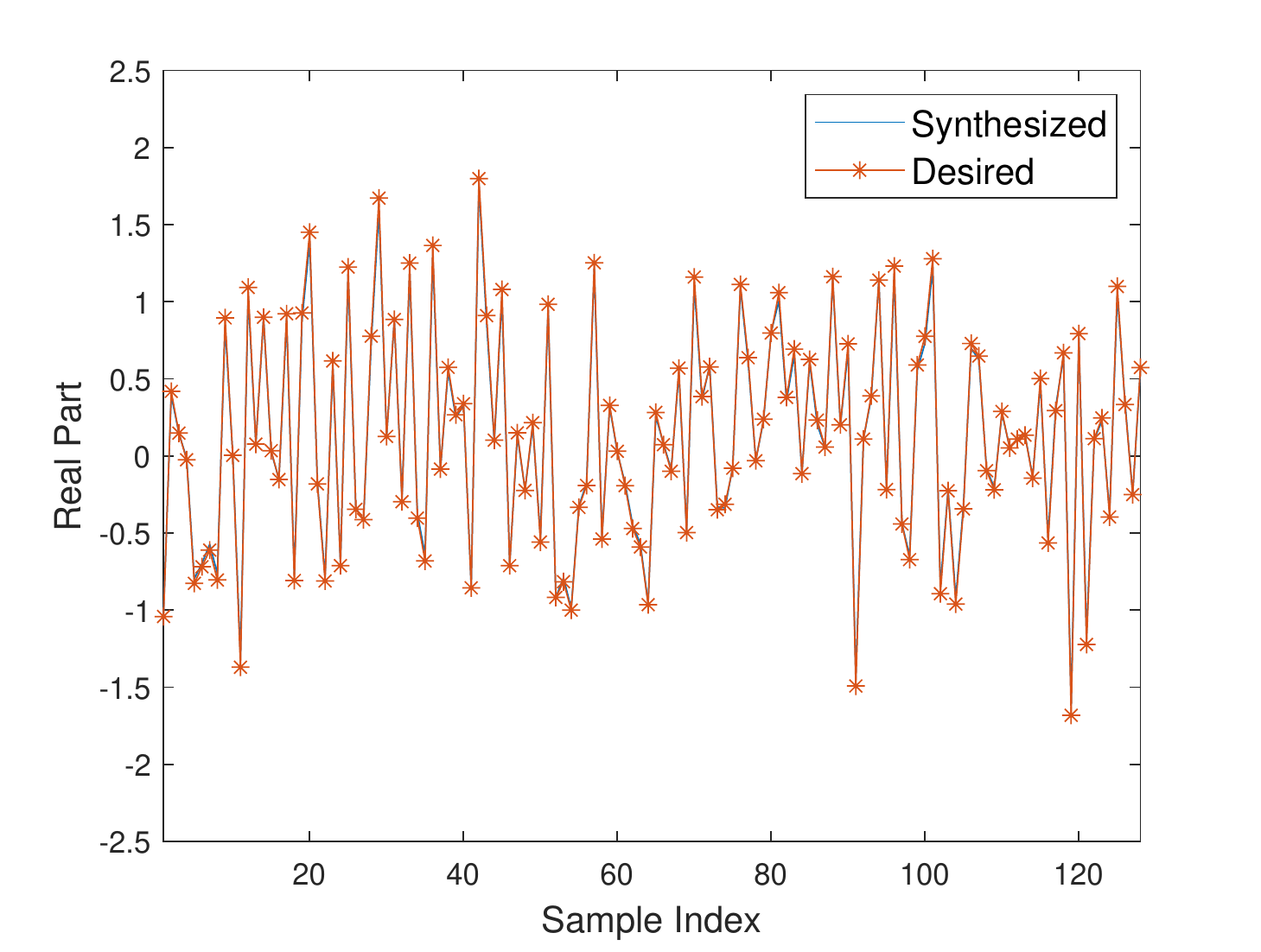}} \label{Fig:8a}} }
{\subfigure[]{{\includegraphics[width = 0.45\textwidth]{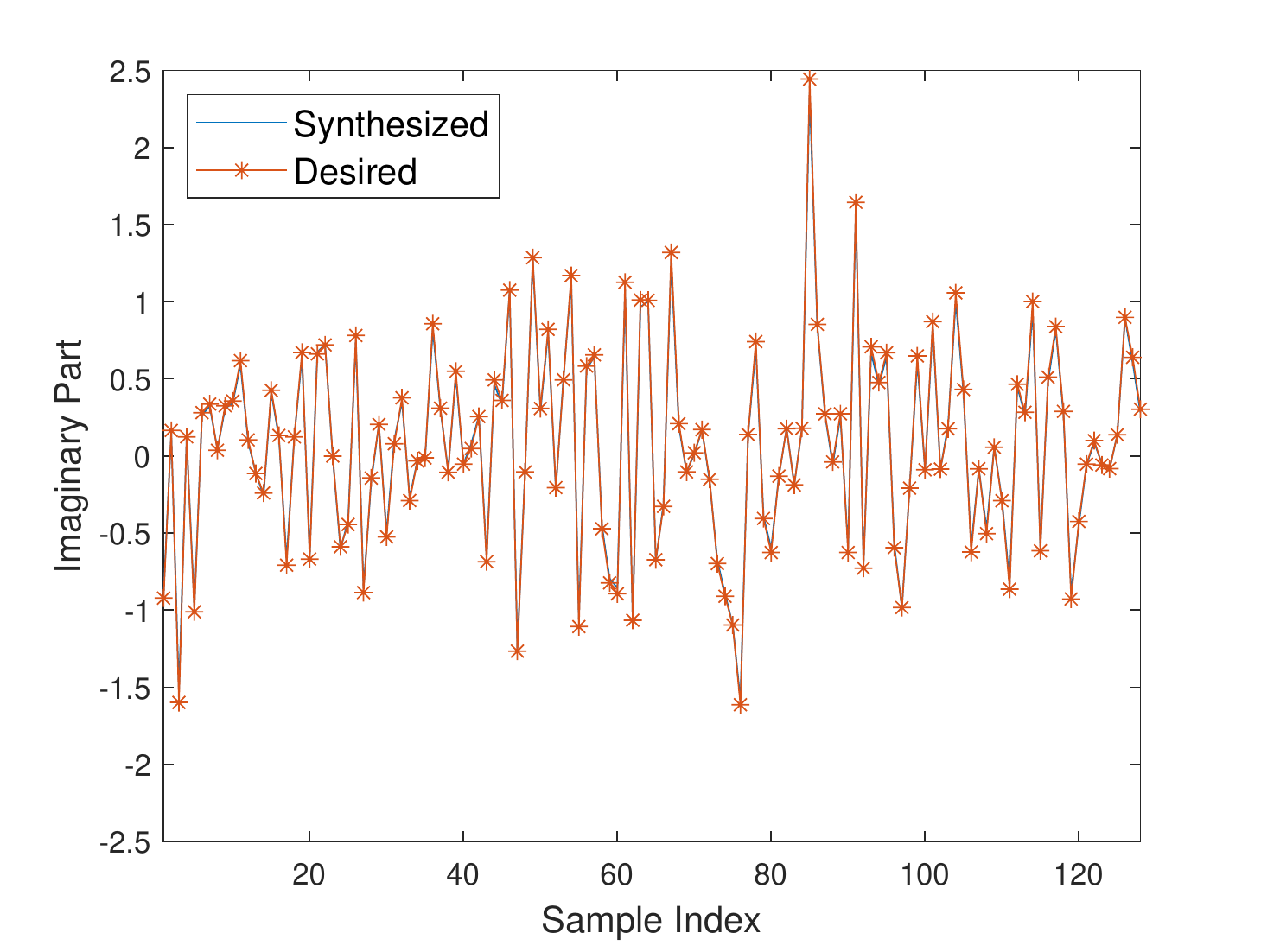}} \label{Fig:8b}} }
\caption{The synthesized noise-like signals at $\theta_{\textrm{jam}} = 20^\circ$.  $\rho = 1$. (a) The real part. (b) The imaginary part. }
\label{Fig:8}
\end{figure}

In this subsection, we analyze the performance of low PAPR waveforms. Fig. \ref{Fig:6} illustrates the convergence of the proposed algorithm for synthesizing constant-modulus waveforms (i.e., waveforms with PAPR $\rho = 1$), when we use the same parameter setting as in Fig. \ref{Fig:2} and Fig. \ref{Fig:3}, except that the transmit energy is $e_t = 500$. In the algorithm, the maximum allowed matching errors are $\varepsilon_1 = 10^{-3}$ and $\varepsilon_2 = 0.2$, and the penalty parameter is set to $\mu = 5$. In addition, we initialize the algorithm with a constant-modulus waveform with modulus equal to $a_s$ and randomly generated phases.
Note the fast convergence of the proposed algorithm. The transmit SINR at convergence is $37.53$ dB. Compared with the radar-only system, whose achievable SINR is $37.78$ dB, the SINR loss is $0.25$ dB (for the energy-constrained waveforms in Fig. \ref{Fig:4d}, the SINR loss is $0.095$ dB).
Fig. \ref{Fig:7} shows the synthesized communication signals for $\theta_\textrm{c} = -25^\circ$. Observe that even for the challenging constant-modulus constraint, the difference between the synthesized signals and the desired signals is small. Moreover, the synthesized signals have a perfect constellation diagram. Fig. \ref{Fig:8} displays the synthesized noise-like jamming signals at  $\theta_{\textrm{jam}} = 20^\circ$. Again, the synthesized signals and the desired signals almost overlap with one another.

\begin{figure}[!htp]
\centering
{\subfigure[]{{\includegraphics[width = 0.45\textwidth]{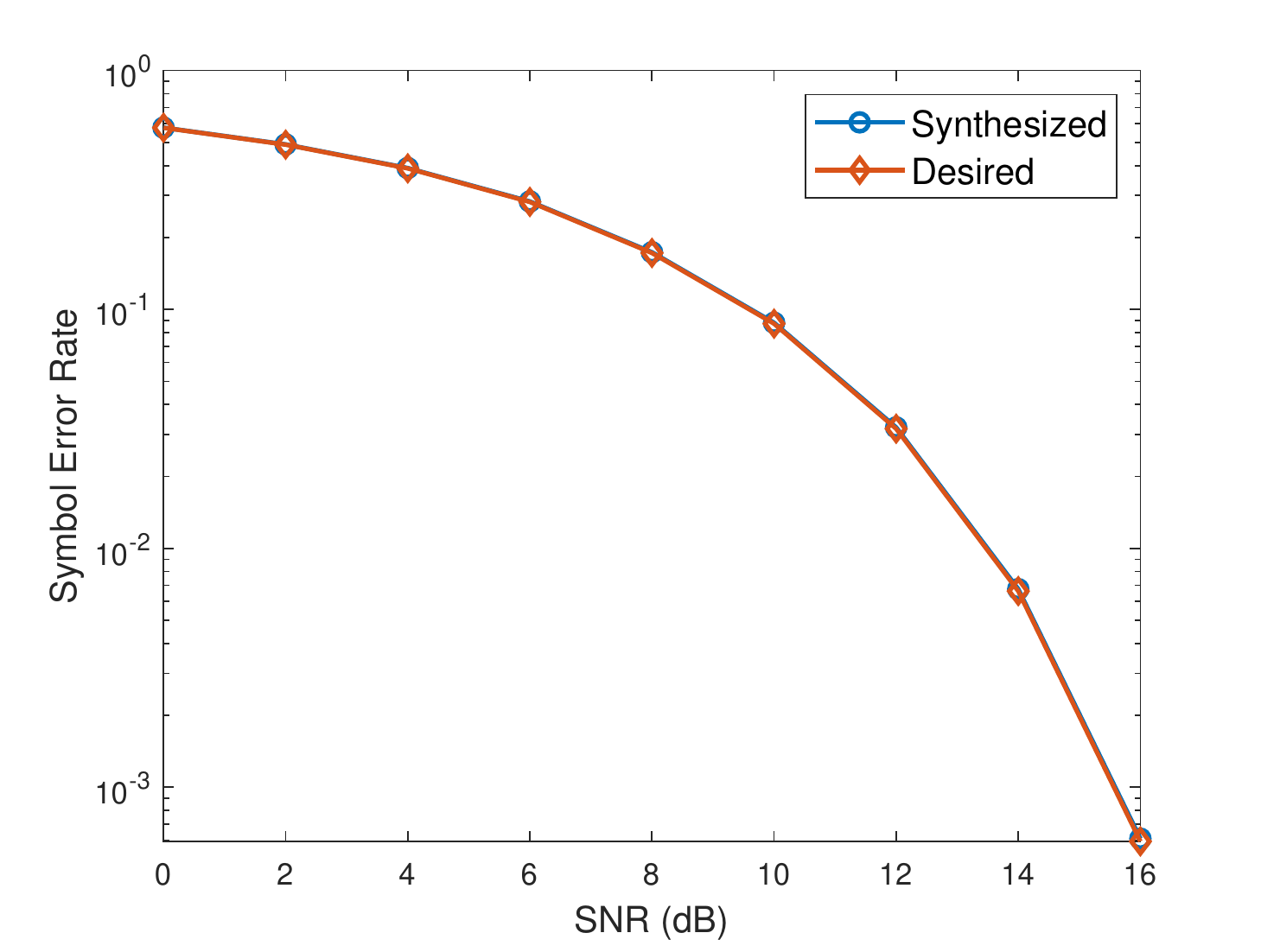}} \label{Fig:9a}} }
{\subfigure[]{{\includegraphics[width = 0.45\textwidth]{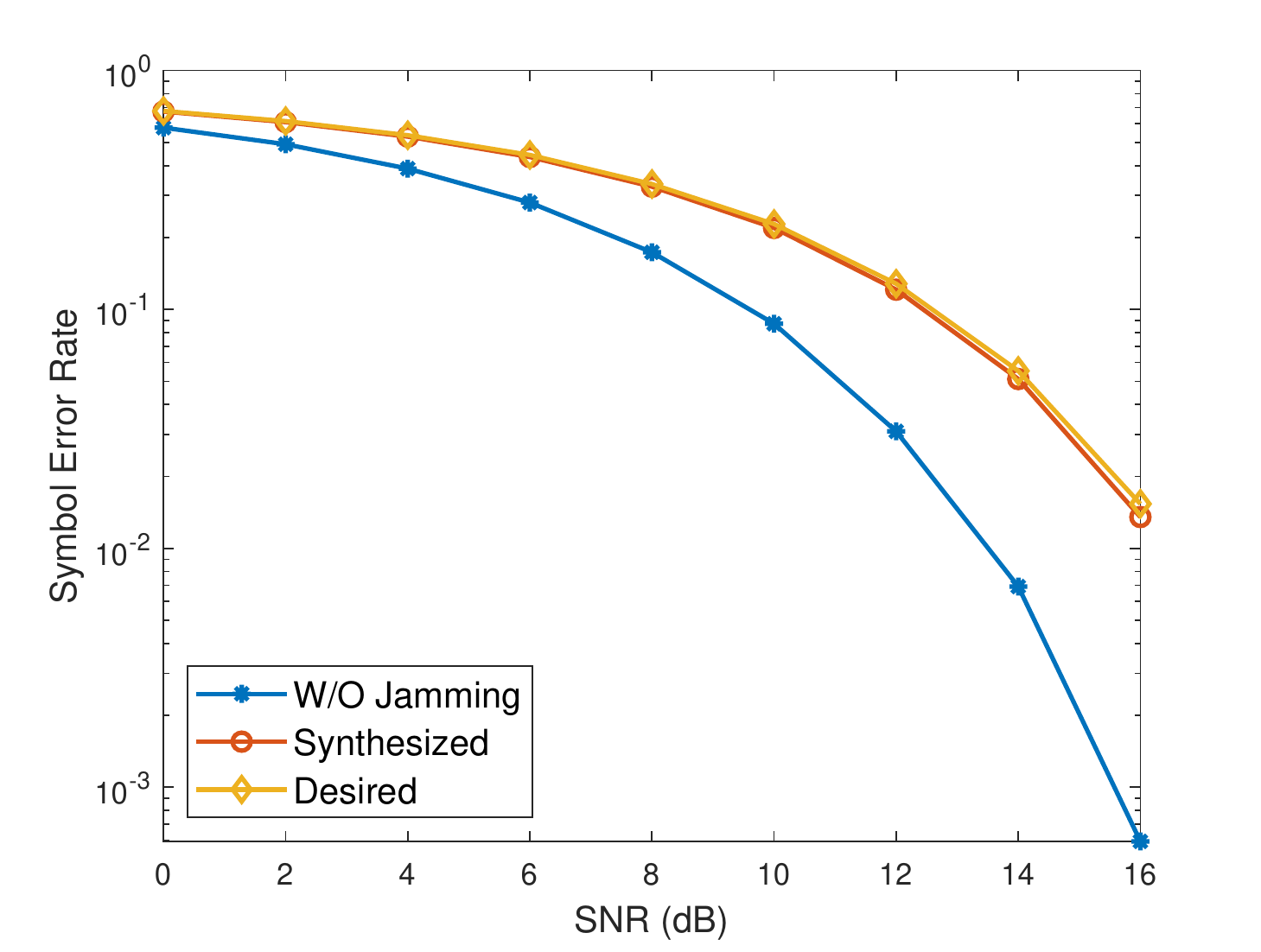}} \label{Fig:9b}} }
\caption{ (a) The symbol error rate of the cooperative communication receiver. (b) The symbol error rate of the hostile target. }
\label{Fig:9}
\end{figure}

Next we assess the performance of the signals received by the (cooperative) communication receiver and the hostile target. We assume that the hostile target is also receiving the 8PSK signals from its peer nodes. Thus, the signals received by the hostile target include both the 8PSK signals and the noise-like signals sent by the MFRF system. Fig. \ref{Fig:9a} shows the symbol error rate (SER) of the communication receiver versus the CSNR, where the SER associated with the desired communication signals is also included as a benchmark; the SERs are obtained from $1000$ independent Monte Carlo trials, and the parameters of the setup are the same as in Fig. \ref{Fig:6}. It can be seen that the performance of the synthesized communication signals is identical to  that of the desired signals. Fig. \ref{Fig:9b} plots the SER associated with the hostile target, when it is jammed by the signals transmitted by the MFRF system, and the jammer-to-noise-ratio is $0$ dB. For comparison, the SER curves associated with the case of sending the desired noise-like signals and the case without jamming are also presented. The jamming signals transmitted by the MFRF system significantly degrade the performance of the hostile target. Moreover, the performance of the transmitted signals is close to that of the desired noise-like signals.

\begin{figure}[!htp]
\centering
\includegraphics[width = 0.45\textwidth]{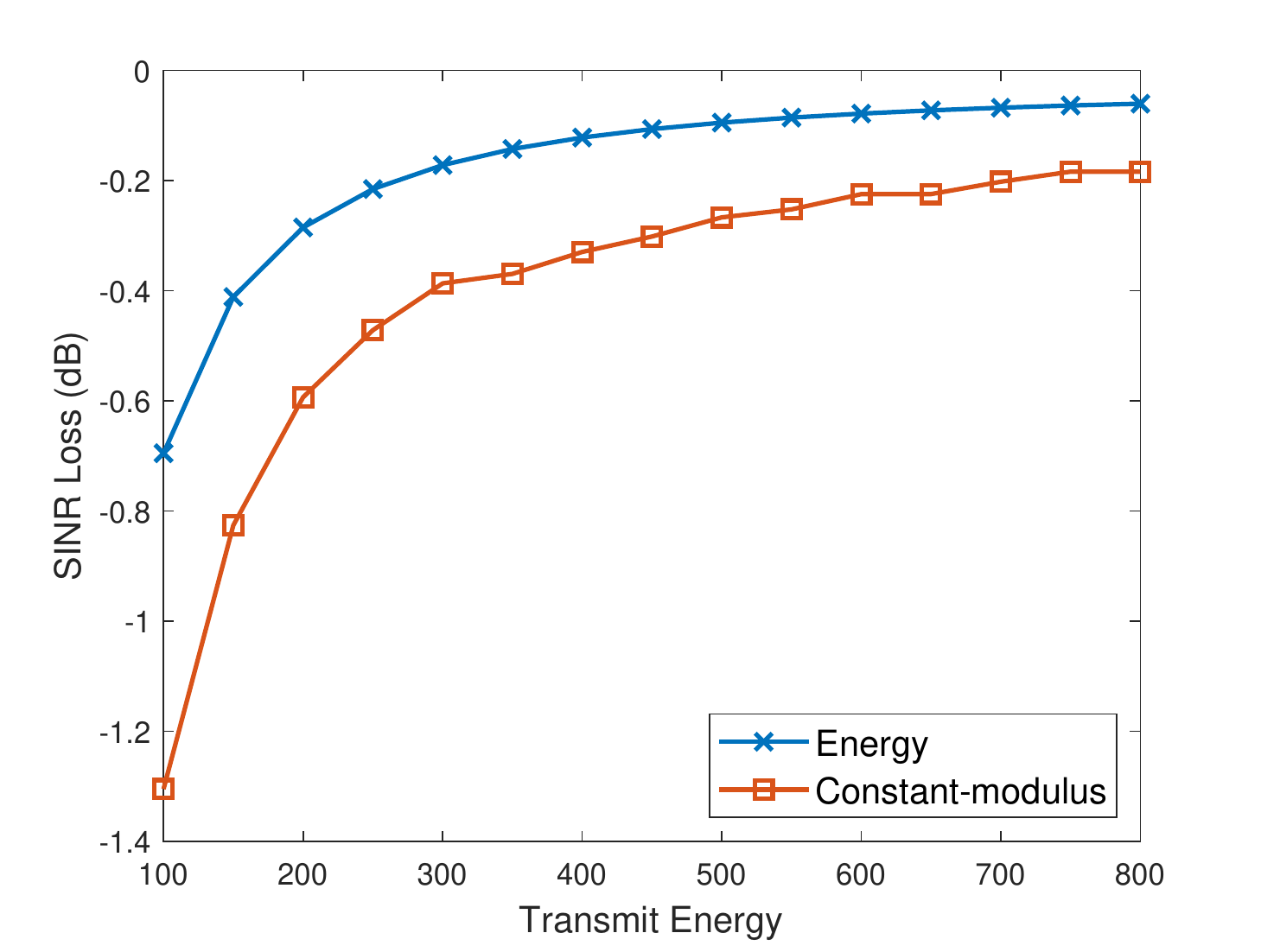}
\caption{The SINR loss of the MFRF system compared with the radar-only system. $L=128$. $\rho = 1$.}
\label{Fig:10}
\end{figure}

To conclude this section,  we compare the SINR of the MFRF system with that of the radar-only system. Fig. \ref{Fig:10} shows the SINR loss of the synthesized constant-modulus waveforms, compared with the radar-only system, versus the transmit energy. The SINR loss associated with the energy-constrained waveforms is also shown as a benchmark. The results indicate that for this parameter setting, if the transmit energy is larger than $150$, the SINR loss will be less than $0.83$ dB, and the performance degradation compared with the energy-constrained waveforms will be smaller than $0.42$ dB.

\section{Conclusions} \label{sec:Conclusion}
This paper has investigated the target detection performance of MIMO MFRF systems. To study the fundamental limits of the detection performance, we formulated an optimization problem for the waveform design and derived the optimal solution. For the case of a structured interference covariance matrix (having a Kronecker product structure), we obtained a simplified solution and analyzed the achievable SINR of the system. Moreover, we studied the system performance when transmitting constant-modulus waveforms. Our results demonstrated that by carefully synthesizing the transmit waveforms, the MIMO MFRF system can provide multiple functions, including radar detection, communication, and jamming of hostile targets. Moreover, the detection performance of the MFRF system can approach that of the radar-only system, if the transmit energy of the desired communication and jamming signals is reasonably small.

Possible topics for future research include the design of low-PAPR waveforms for MIMO MFRF systems in the presence of clutter; robust waveform design in the presence of angle estimation errors (i.e., the direction estimates of the target, communication receivers, and the hostile targets are inaccurate); derivation of computationally efficient algorithms that would allow the MFRF systems to optimize the waveforms in real time; and investigation of information-theoretic bounds for MIMO MFRF systems (see, e.g., \cite{Bliss2014bound, Chiriyath2016Bounds} for some discussions on this topic).

\appendices

%
\section{Derivation of \eqref{eq:CoherentWaveform}}
For the case of $N_0=1$, $\hat{\bS} = \NT^{-1} \ba(\bar{\theta}) \bd^{\top}$.
Moreover, $\bB\bB^\HH$ can be written as
\begin{equation}
  \bB\bB^\HH = \bI - \NT^{-1}\ba(\bar{\theta})\ba^\HH(\bar{\theta}).
\end{equation}
Then, after some algebraic manipulations, we can verify that
\begin{align}\label{eq:ApA2}
  \bB\bB^\HH \ba(\theta_\textrm{t})\ba^\HH (\theta_\textrm{t})\hat{\bS}
  &=[\ba(\theta_\textrm{t}) - B(\theta_\textrm{t},\bar{\theta})\ba(\bar{\theta})]\cdot B^*(\theta_\textrm{t},\bar{\theta})\bd^{\top} \nonumber \\
  &=\ba(\theta_\textrm{t})\bd^{\top} - G^2(\theta_\textrm{t},\bar{\theta})\ba(\bar{\theta})\bd^{\top}.
\end{align}
Substituting the result in \eqref{eq:ApA2} and the expression for $\hat{\bS}$ into \eqref{eq:WaveformUnderEnergyConstraint}, we obtain \eqref{eq:CoherentWaveform}.

%
%

\ifCLASSOPTIONcaptionsoff
  \newpage
\fi

\bibliographystyle{IEEEtran}
\bibliography{IEEEabrv,robustMFRF}

\begin{thebibliography}{10}
\providecommand{\url}[1]{#1}
\csname url@samestyle\endcsname
\providecommand{\newblock}{\relax}
\providecommand{\bibinfo}[2]{#2}
\providecommand{\BIBentrySTDinterwordspacing}{\spaceskip=0pt\relax}
\providecommand{\BIBentryALTinterwordstretchfactor}{4}
\providecommand{\BIBentryALTinterwordspacing}{\spaceskip=\fontdimen2\font plus
\BIBentryALTinterwordstretchfactor\fontdimen3\font minus
  \fontdimen4\font\relax}
\providecommand{\BIBforeignlanguage}[2]{{%
\expandafter\ifx\csname l@#1\endcsname\relax
\typeout{** WARNING: IEEEtran.bst: No hyphenation pattern has been}%
\typeout{** loaded for the language `#1'. Using the pattern for}%
\typeout{** the default language instead.}%
\else
\language=\csname l@#1\endcsname
\fi
#2}}
\providecommand{\BIBdecl}{\relax}
\BIBdecl

\bibitem{Tang2021radarconf}
B.~Tang, Z.~Huang, L.~Qin, and H.~Wang, ``Fundamental limits on detection with
  a dual-function radar communication system,'' in \emph{CIE International
  Conference on Radar}, 2021, pp. 1--5.

\bibitem{tavik2005advanced}
G.~C. Tavik, C.~L. Hilterbrick, J.~B. Evins, J.~J. Alter, J.~G. Crnkovich,
  J.~W. de~Graaf, W.~Habicht, G.~P. Hrin, S.~A. Lessin, D.~C. Wu \emph{et~al.},
  ``The advanced multifunction {RF} concept,'' \emph{IEEE Transactions on
  Microwave Theory and Techniques}, vol.~53, no.~3, pp. 1009--1020, 2005.

\bibitem{moo2018overview}
P.~W. Moo and D.~J. DiFilippo, ``Overview of naval multifunction {RF}
  systems,'' in \emph{2018 15th European Radar Conference (EuRAD)}.\hskip 1em
  plus 0.5em minus 0.4em\relax IEEE, 2018, pp. 178--181.

\bibitem{hemmi1996ASAP}
C.~Hemmi, R.~T. Dover, A.~Vespa, and M.-W. Fenton, ``Advanced shared aperture
  program ({ASAP}) array design,'' in \emph{Proceedings of International
  Symposium on Phased Array Systems and Technology}.\hskip 1em plus 0.5em minus
  0.4em\relax IEEE, 1996, pp. 278--282.

\bibitem{ouacha2010MAESA}
A.~Ouacha, A.~Fredlund, J.~Andersson, H.~Hindsefelt, V.~Rinaldi, and
  C.~Scattoni, ``{SE-IT} joint {M-AESA} program: Overview and status,'' in
  \emph{2010 IEEE International Symposium on Phased Array Systems and
  Technology}.\hskip 1em plus 0.5em minus 0.4em\relax IEEE, 2010, pp. 771--776.

\bibitem{Mealey1963RadarCom}
R.~M. Mealey, ``A method for calculating error probabilities in a radar
  communication system,'' \emph{IEEE Transactions on Space Electronics and
  Telemetry}, vol.~9, no.~2, pp. 37--42, 1963.

\bibitem{barrenechea2007fmcw}
P.~Barrenechea, F.~Elferink, and J.~Janssen, ``{FMCW} radar with broadband
  communication capability,'' in \emph{2007 European Radar Conference}.\hskip
  1em plus 0.5em minus 0.4em\relax IEEE, 2007, pp. 130--133.

\bibitem{Sturm2011OFDM}
C.~Sturm and W.~Wiesbeck, ``Waveform design and signal processing aspects for
  fusion of wireless communications and radar sensing,'' \emph{Proceedings of
  the IEEE}, vol.~99, no.~7, pp. 1236--1259, 2011.

\bibitem{Liu2017OFDM}
Y.~Liu, G.~Liao, Z.~Yang, and J.~Xu, ``Multiobjective optimal waveform design
  for {OFDM} integrated radar and communication systems,'' \emph{Signal
  Processing}, vol. 141, pp. 331--342, 2017.

\bibitem{chen2011LFMMSK}
X.~Chen, X.~Wang, S.~Xu, and J.~Zhang, ``A novel radar waveform compatible with
  communication,'' in \emph{2011 International Conference on Computational
  Problem-Solving (ICCP)}.\hskip 1em plus 0.5em minus 0.4em\relax IEEE, 2011,
  pp. 177--181.

\bibitem{nowak2017mixed}
M.~J. Nowak, Z.~Zhang, L.~LoMonte, M.~Wicks, and Z.~Wu, ``Mixed-modulated
  linear frequency modulated radar-communications,'' \emph{IET Radar, Sonar \&
  Navigation}, vol.~11, no.~2, pp. 313--320, 2017.

\bibitem{Ciuonzo2015Intrapulse}
D.~Ciuonzo, A.~De~Maio, G.~Foglia, and M.~Piezzo, ``Intrapulse radar-embedded
  communications via multiobjective optimization,'' \emph{IEEE Transactions on
  Aerospace and Electronic Systems}, vol.~51, no.~4, pp. 2960--2974, 2015.

\bibitem{Hassanien2016Embedding}
A.~Hassanien, M.~G. Amin, Y.~D. Zhang, and F.~Ahmad, ``Dual-function
  radar-communications: Information embedding using sidelobe control and
  waveform diversity,'' \emph{IEEE Transactions on Signal Processing}, vol.~64,
  no.~8, pp. 2168--2181, 2016.

\bibitem{Wang2019SparseArray}
X.~Wang, A.~Hassanien, and M.~G. Amin, ``Dual-function {MIMO} radar
  communications system design via sparse array optimization,'' \emph{IEEE
  Transactions on Aerospace and Electronic Systems}, vol.~55, no.~3, pp.
  1213--1226, 2019.

\bibitem{McCormick2017Simultaneous}
P.~M. McCormick, S.~D. Blunt, and J.~G. Metcalf, ``Simultaneous radar and
  communications emissions from a common aperture, part i: Theory,'' in
  \emph{IEEE Radar Conference (RadarConf)}, 2017, Conference Proceedings, pp.
  1685--1690.

\bibitem{Liu2018DFRC}
F.~Liu, L.~Zhou, C.~Masouros, A.~Li, W.~Luo, and A.~Petropulu, ``Toward
  dual-functional radar-communication systems: Optimal waveform design,''
  \emph{IEEE Transactions on Signal Processing}, vol.~66, no.~16, pp.
  4264--4279, 2018.

\bibitem{Zhang2021Overview}
J.~A. Zhang, F.~Liu, C.~Masouros, R.~W. Heath, Z.~Feng, L.~Zheng, and
  A.~Petropulu, ``An overview of signal processing techniques for joint
  communication and radar sensing,'' \emph{IEEE Journal of Selected Topics in
  Signal Processing}, vol.~15, no.~6, pp. 1295--1315, 2021.

\bibitem{Liu2020Beamforming}
X.~Liu, T.~Huang, N.~Shlezinger, Y.~Liu, J.~Zhou, and Y.~C. Eldar, ``Joint
  transmit beamforming for multiuser {MIMO} communications and {MIMO} radar,''
  \emph{IEEE Transactions on Signal Processing}, vol.~68, pp. 3929--3944, 2020.

\bibitem{Tang2020SAM}
B.~Tang, H.~Wang, L.~Qin, and L.~Li, ``Waveform design for dual-function {MIMO}
  radar-communication systems,'' in \emph{IEEE 11th Sensor Array and
  Multichannel Signal Processing Workshop (SAM)}, 2020, Conference Proceedings,
  pp. 1--5.

\bibitem{Shi2020DFRC}
S.~Shi, Z.~Wang, Z.~He, and Z.~Cheng, ``Constrained waveform design for
  dual-functional {MIMO} radar-communication system,'' \emph{Signal
  Processing}, vol. 171, p. 107530, 2020.

\bibitem{Tsinos2020JointDesign}
C.~G. Tsinos, A.~Arora, S.~Chatzinotas, and B.~Ottersten, ``Joint transmit
  waveform and receive filter design for dual-function radar-communication
  systems,'' \emph{IEEE Journal of Selected Topics in Signal Processing},
  vol.~15, no.~6, pp. 1378--1392, 2021.

\bibitem{li2022dual}
Y.~Li and A.~Petropulu, ``Dual-function radar-communication system aided by
  intelligent reflecting surfaces,'' \emph{arXiv preprint arXiv:2204.04721},
  2022.

\bibitem{Li2007mimoIntroduction}
J.~Li and P.~Stoica, ``{MIMO} radar with colocated antennas,'' \emph{{IEEE}
  Signal Process. Mag.}, vol.~24, no.~5, pp. 106--114, 2007.

\bibitem{Xu2008MIMO}
L.~Xu, J.~Li, and P.~Stoica, ``Target detection and parameter estimation for
  {MIMO} radar systems,'' \emph{IEEE Transactions on Aerospace and Electronic
  Systems}, vol.~44, no.~3, pp. 927--939, 2008, 0018-9251.

\bibitem{Roberts2010IAA}
W.~Roberts, P.~Stoica, J.~Li, T.~Yardibi, and F.~A. Sadjadi, ``Iterative
  adaptive approaches to {MIMO} radar imaging,'' \emph{IEEE Journal of Selected
  Topics in Signal Processing}, vol.~4, no.~1, pp. 5--20, 2010.

\bibitem{Stoica2007Probing}
P.~Stoica, J.~Li, and Y.~Xie, ``On probing signal design for {MIMO} radar,''
  \emph{IEEE Transactions on Signal Processing}, vol.~55, no.~8, pp.
  4151--4161, 2007.

\bibitem{HornJohnson1990matrixbook}
R.~A. Horn and C.~R. Johnson, \emph{Matrix analysis}.\hskip 1em plus 0.5em
  minus 0.4em\relax Cambridge: Cambridge University Press, 1990.

\bibitem{kaybook1998}
S.~M. Kay, \emph{Fundamentals of Statistical Signal Processing-Volume {II}:
  Detection Theory}.\hskip 1em plus 0.5em minus 0.4em\relax New Jersey:
  Prentice Hall, 1998.

\bibitem{Larsson2013precoding}
S.~K. Mohammed and E.~G. Larsson, ``Per-antenna constant envelope precoding for
  large multi-user {MIMO} systems,'' \emph{IEEE Transactions on
  Communications}, vol.~61, no.~3, pp. 1059--1071, 2013.

\bibitem{stoica2005spectral}
P.~Stoica and R.~Moses, \emph{Spectral analysis of signals}.\hskip 1em plus
  0.5em minus 0.4em\relax Upper Saddle River, NJ: Prentice Hall, 2005.

\bibitem{Huang2007decomposition}
Y.~Huang and S.~Zhang, ``Complex matrix decomposition and quadratic
  programming,'' \emph{Mathematics of Operations Research}, vol.~32, no.~3, pp.
  758--768, 2007.

\bibitem{Liu2015NoTraining}
W.~Liu, Y.~Wang, J.~Liu, W.~Xie, H.~Chen, and W.~Gu, ``Adaptive detection
  without training data in colocated {MIMO} radar,'' \emph{IEEE Transactions on
  Aerospace and Electronic Systems}, vol.~51, no.~3, pp. 2469--2479, 2015.

\bibitem{Liu2018Tunable}
J.~Liu, S.~Zhou, W.~Liu, J.~Zheng, H.~Liu, and J.~Li, ``Tunable adaptive
  detection in colocated {MIMO} radar,'' \emph{IEEE Transactions on Signal
  Processing}, vol.~66, no.~4, pp. 1080--1092, 2018.

\bibitem{bernstein2009matrix}
D.~S. Bernstein, \emph{Matrix mathematics: theory, facts, and formulas}.\hskip
  1em plus 0.5em minus 0.4em\relax Princeton University Press, 2009.

\bibitem{li2010phased}
J.~Li and P.~Stoica, ``The phased array is the maximum {SNR} active array,''
  \emph{IEEE Signal Processing Magazine}, vol.~27, no.~2, pp. 143--144, 2010.

\bibitem{Stoica1989CRB}
P.~Stoica and A.~Nehorai, ``{MUSIC}, maximum likelihood, and {Cramer-Rao}
  bound,'' \emph{IEEE Transactions on Acoustics, Speech and Signal Processing},
  vol.~37, no.~5, pp. 720--741, 1989.

\bibitem{Boyd2011ADMM}
S.~Boyd, N.~Parikh, E.~Chu, B.~Peleato, and J.~Eckstein, ``Distributed
  optimization and statistical learning via the alternating direction method of
  multipliers,'' \emph{Foundations and Trends® in Machine Learning}, vol.~3,
  no.~1, pp. 1--122, 2011.

\bibitem{Tang2021IT}
B.~Tang and P.~Stoica, ``Information-theoretic waveform design for {MIMO} radar
  detection in range-spread clutter,'' \emph{Signal Processing}, vol. 182, p.
  107961, 2021.

\bibitem{Tang2021Profiling}
B.~Tang, J.~Liu, H.~Wang, and Y.~Hu, ``Constrained radar waveform design for
  range profiling,'' \emph{IEEE Transactions on Signal Processing}, vol.~69,
  pp. 1924--1937, 2021.

\bibitem{Tropp2005AP}
J.~A. Tropp, I.~S. Dhillon, R.~W. Heath, and T.~Strohmer, ``Designing
  structured tight frames via an alternating projection method,'' \emph{IEEE
  Transactions on information theory}, vol.~51, no.~1, pp. 188--209, 2005.

\bibitem{Beck2009FISTA}
A.~Beck and M.~Teboulle, ``A fast iterative shrinkage-thresholding algorithm
  for linear inverse problems,'' \emph{SIAM journal on imaging sciences},
  vol.~2, no.~1, pp. 183--202, 2009.

\bibitem{varadhan2008SQUAREM}
R.~Varadhan and C.~Roland, ``Simple and globally convergent methods for
  accelerating the convergence of any {EM} algorithm,'' \emph{Scandinavian
  Journal of Statistics}, vol.~35, no.~2, pp. 335--353, 2008.

\bibitem{Bliss2014bound}
D.~W. Bliss, ``Cooperative radar and communications signaling: The estimation
  and information theory odd couple,'' in \emph{IEEE Radar Conference}.\hskip
  1em plus 0.5em minus 0.4em\relax IEEE, 2014, Conference Proceedings, pp.
  50--55.

\bibitem{Chiriyath2016Bounds}
A.~R. Chiriyath, B.~Paul, G.~M. Jacyna, and D.~W. Bliss, ``Inner bounds on
  performance of radar and communications co-existence,'' \emph{IEEE
  Transactions on Signal Processing}, vol.~64, no.~2, pp. 464--474, 2016.

\end{thebibliography}

%
%


\end{document}